\documentclass[12pt]{article}
\usepackage{amsmath,amsfonts,amssymb,graphics,psfrag}
\usepackage{mathrsfs, dsfont}
\usepackage{multirow,array,epsfig,multirow,stmaryrd,graphicx}
\usepackage{comment}
\usepackage{sectsty}
\usepackage[letterpaper,top=1in,bottom=1.15in,left=1in,right=1in]{geometry}
\usepackage[format=hang,margin=1.5cm,font=small,labelfont=bf]{caption}
\usepackage{subfigure}

\usepackage[dvips,bookmarks,breaklinks]{hyperref}		
\usepackage[all,cmtip]{xy}

\numberwithin{equation}{section}
\numberwithin{table}{section}

\sectionfont{\large\bfseries}
\subsectionfont{\normalsize\bfseries}
\subsubsectionfont{\normalsize\it}

\newcommand{\beq}{\begin{equation}}
\newcommand{\eeq}{\end{equation}}
\newcommand{\be}{\begin{equation}} 
\newcommand{\ee}{\end{equation}}
\newcommand{\bea}{\begin{eqnarray}}
\newcommand{\eea}{\end{eqnarray}}   
\newcommand{\ben}{\begin{eqnarray*}}
\newcommand{\een}{\end{eqnarray*}}                  
\newcommand{\ba}{\begin{aligned}}
\newcommand{\ea}{\end{aligned}}
\newcommand{\bt}{\begin{tabular}}
\newcommand{\et}{\end{tabular}}
\newcommand{\bc}{\begin{center}}
\newcommand{\ec}{\end{center}}

\newcommand{\cC}{\mathcal{C}}

\newcommand{\cL}{\mathcal{L}}

\newcommand{\cN}{\mathcal{N}}

\newcommand{\cF}{\mathcal{F}}

\newcommand{\cR}{\mathcal{R}}

\newcommand{\nn}{\nonumber}

\newcommand{\cref}{{\bf [check ref]}}

\newcommand{\tr}{\mathrm{Tr}\:}

\psfrag{n1}{$\nu_1$}
\psfrag{n2}{$\nu_2$}
\psfrag{n1'}{$\nu_1'$}
\psfrag{n2'}{$\nu_2'$}
\psfrag{n9}{$\nu_9$}
\psfrag{n10'}{$\nu_{10}'$}
\psfrag{t1}{$\tilde{\nu}_1$}
\psfrag{t2}{$\tilde{\nu}_2$}
\psfrag{t9}{$\tilde{\nu}_9$}
\psfrag{t1'}{$\tilde{\nu}_1'$}
\psfrag{t2'}{$\tilde{\nu}_2'$}
\psfrag{t10'}{$\tilde{\nu}_{10}'$}

\entrymodifiers={+!!<0pt,\fontdimen22\textfont2>}		

\def\Z{\mathds{Z}}

\def\C{\mathds{C}}

\def\P{\mathds{P}}

\newcommand\T{\rule{0pt}{2.6ex}}			
\newcommand\B{\rule[-1.2ex]{0pt}{0pt}}		
\numberwithin{equation}{section}			

\makeatletter
\renewcommand*\env@matrix[1][*\c@MaxMatrixCols c]{%
  \hskip -\arraycolsep
  \let\@ifnextchar\new@ifnextchar
  \array{#1}}
\makeatother

\begin{document}

\baselineskip=17pt

\begin{titlepage}
\begin{flushright}
\parbox[t]{1.8in}{
BONN-TH-2009-07\\
0912.2997\ [hep-th]}
\end{flushright}

\begin{center}

\vspace*{ 1.2cm}

{\large \bf Five-Brane Superpotentials and Heterotic/F-theory Duality
}

\vskip 1.2cm

\begin{center}
 \bf{Thomas W.~Grimm, Tae-Won Ha, Albrecht Klemm and Denis Klevers} \footnote{\texttt{grimm, tha, aklemm, klevers@th.physik.uni-bonn.de}}
\end{center}
\vskip .2cm

{\em Bethe Center for Theoretical Physics, Universit\"at Bonn, \\[.1cm]
Nussallee 12, 53115 Bonn, Germany}
 \vspace*{1cm}

\end{center}

\vskip 0.2cm

\begin{center} {\bf ABSTRACT } \end{center}

Under heterotic/F-theory duality it was argued that a wide class of heterotic
five-branes is mapped into the geometry of an F-theory compactification
manifold. In four-dimensional compactifications this identifies a five-brane
wrapped on a curve in the base of an elliptically fibered Calabi-Yau threefold
with a specific F-theory Calabi-Yau fourfold containing the blow-up of the
five-brane curve. We argue that this duality can be reformulated by first
constructing a non-Calabi-Yau heterotic threefold by blowing up the curve of
the five-brane into a divisor with five-brane flux.  Employing
heterotic/F-theory duality this leads us to the construction of a Calabi-Yau
fourfold and four-form flux. Moreover, we obtain an explicit map between the
five-brane superpotential and an F-theory flux superpotential.  The map of the
open-closed deformation problem of a five-brane in a compact Calabi-Yau
threefold into a deformation problem of complex structures on a dual Calabi-Yau
fourfold with four-form flux  provides a powerful tool to explicitly compute
the five-brane superpotential.

\hfill December, 2009
\end{titlepage}

\section{Introduction}

The study of string compactifications leading to $\cN=1$ supersymmetric
four-dimensional low-energy effective theories is of conceptual as well as of
phenomenological interest.  Two prominent approaches to obtain such effective
theories are  either to consider heterotic $E_8 \times E_8$ string theory on a
Calabi-Yau manifold with non-trivial vector bundles, or to study F-theory
compactifications on singular Calabi-Yau fourfolds.  At first, these two
approaches appear to be very different in nature, since the data determining the
effective dynamics are encoded by seemingly different objects.  However, at
least if one focuses on certain compact geometries, the heterotic and F-theory
picture are believed to be dual descriptions of the same physics
\cite{Vafa:1996xn,Morrison:1996na}.  The dictionary of this duality not only
contains the map for the vector bundles of the heterotic string, but also
includes heterotic five-branes wrapped on curves in the Calabi-Yau threefold
\cite{Andreas:1997ce,Berglund:1998ej,Rajesh:1998ik}. These are often necessary
in a consistent heterotic compactification to ensure anomaly cancellation.
Using this duality either of the two descriptions can be used to answer specific
questions about the four-dimensional physics. In this work we will focus on
parts of the effective action which are efficiently calculable in F-theory, but
admit a natural physical interpretation in the heterotic theory.

An important question in the study of the four-dimensional $\cN=1$ low-energy
effective action is the explicit computation of the superpotential and
gauge-coupling functions which depend holomorphically on the chiral multiplets.
In the following we will mainly focus on the study of the superpotential of
a heterotic five-brane wrapped on a curve $\cC$ in a Calabi-Yau threefold
$Z_3$. It was shown in ref.~\cite{Witten:1997ep} that this superpotential
depends on the deformation modes of the curve $\cC$ and the complex structure
moduli of $Z_3$ via the chain integral $\int_{\Gamma} \Omega$, where $\Omega$
is the holomorphic three-form on $Z_3$, and $\Gamma$ is a three-chain which
admits $\cC$ as a boundary component. We will argue by using heterotic/F-theory
duality that this chain integral is mapped to the flux superpotential of an
F-theory compactification upon constructing an appropriate Calabi-Yau fourfold
$\hat X_4$ encoding the five-brane dynamics, and the associated four-form flux
$G_4$.  The F-theory flux superpotential can then be computed explicitly by
solving Picard-Fuchs differential equations determining the closed period
integrals of the holomorphic four-form on $\hat X_4$, and using mirror symmetry
to identify the superpotential solution \cite{Grimm:2009ef}. Earlier
discussions and computations of the periods of the holomorphic four-form can be
found in refs.~\cite{Greene:1993vm,Mayr:1996sh,Klemm:1996ts}.

The computation of brane superpotentials given by chain integrals has been of
significant interest in the D-brane literature.  Starting with
\cite{Aganagic:2000gs} the superpotential for D5-branes has been studied
intensively for non-compact Calabi-Yau threefolds
\cite{Aganagic:2001nx,Lerche:2002ck}.  More recently, there has been various
attempts to extend this to compact threefolds
\cite{Walcher,KSch,Jockers,Baumgartl:2007an,
Grimm:2008dq,Alim:2009rf,Alim:2009bx, Li:2009dz}.  In particular, it was
proposed in refs.~\cite{Lerche:2002ck} to use the variation of mixed
Hodge structure for an auxiliary divisor capturing the variations of the curve
$\cC$. While initially studied in non-compact geometries, extensions to compact
Calabi-Yau threefolds with D5-branes have appeared in
refs.~\cite{Jockers,Alim:2009rf,Alim:2009bx}.  In this proposal the
deformations of an appropriately chosen divisor are effectively identified with    
the deformations of the curve. In contrast, it was argued in
\cite{Grimm:2008dq} that the deformation problem of the curve $\cC$ in $Z_3$
admits a natural map to a geometric setup in which the curve is blown up into
a divisor. In this case, the divisor is rigid\footnote{More accurately, such a divisor is described as an isolated divisor.} and the deformations of the curve
appear as new complex structure deformations of the blown-up threefold $\hat
Z_3$. In this work we will use the blow-up construction to study the duality of
an heterotic five-brane to an F-theory compactification fourfold. Let us note that in
refs.~\cite{Lerche:2001cw, Alim:2009rf,Alim:2009bx,Aganagic:2009jq} it was
proposed to use non-compact Calabi-Yau fourfolds to compute the D5-brane
superpotential and a connection with F-theory was indicated.  Let us stress
that the approach we are using here is different in nature, and rather
completes the approach initiated in our works~\cite{Grimm:2008dq,Grimm:2009ef}.

The map of heterotic string theory on a Calabi-Yau threefold $Z_3$ with
five-branes to an F-theory compactification is best studied for elliptically
fibered $Z_3$.  It was shown in ref.~\cite{Friedman:1997yq} that there exist
elegant constructions of heterotic vector bundles on these threefolds.
Furthermore, a five-brane wrapped on a curve $\cC$ in the base $B_2$ of this
elliptic fibration was argued to map entirely into the geometry of an F-theory
compactification. Using the adiabatic argument of \cite{Vafa:1995gm}
the heterotic string on $Z_3$ is equivalent to F-theory on an elliptic K3-
fibered Calabi-Yau fourfold $X_4$ with base $B_2$. This implies, in particular,
that the three-dimensional base $B_3$ of the elliptically fibered F-theory
fourfold $X_4$ is a holomorphic $\mathds{P}^1$-fibration over $B_2$.  It was
then argued in refs.~\cite{Berglund:1998ej,Rajesh:1998ik,Diaconescu:1999it}
that in the presence of a heterotic five-brane one has to blow up the curve
$\cC$ into a rigid divisor in $B_3$.  The deformations of the curve $\cC$ then map to
complex structure deformations of the  blown-up Calabi-Yau fourfold $\hat X_4$,
and hence can be constrained by a calculable flux superpotential. Note that certain
five-branes can also be interpreted as special gauge bundle configurations of the heterotic string,
the so-called small instantons. In
the small instanton/five-brane transition the deformation moduli of the curve
$\cC$ are identified with heterotic bundle moduli.  This yields yet another
identification of superpotentials, since the five-brane superpotential arises
as a localization of the Chern-Simons superpotential for the bundle moduli
\cite{Aganagic:2000gs}. Both types of superpotentials are efficiently calculable
on the F-theory side using the geometric tools for Calabi-Yau fourfolds.

To study the duality map between the heterotic and F-theory setup, one can
alternatively start by blowing up the heterotic threefold $Z_3$ along the
five-brane curve $\cC$ into $\hat Z_3$ \cite{Grimm:2008dq}.  This can be made explicit by
realizing $\hat Z_3$ as a complete intersection. The non-Calabi-Yau threefold
$\hat{Z}_3$ contains the five-brane moduli as a subsector of its complex
structure moduli.  The heterotic superpotential crucially depends  on the
pull-back $\hat \Omega$ of the holomorphic three-form $\Omega$ to $\hat Z_3$.
Since $\hat \Omega$ vanishes along the blow-up divisor $D$, the heterotic flux,    
specifying the five-brane, localizes on elements in $H^{3}(\hat Z_3-D,\Z)$.
This is equivalent to considering relative three-forms in $H^{3}(\hat
Z_3,D,\Z)$.\footnote{This should be compared with the use of relative
cohomology for the auxiliary non-rigid divisor in the constructions of
refs.~\cite{Lerche:2002ck,Jockers,Alim:2009rf,Alim:2009bx}.} Identifying the
elements of this relative group with elements in the dual fourfold cohomology,
one finds an explicit map between the heterotic five-brane and F-theory fluxes.
We propose, and explicitly demonstrate for examples, that the F-theory geometry
$\hat X_4$ can in turn be entirely constructed from $\hat{Z}_3$. In particular, this
identification becomes apparent when also realizing $\hat X_4$ as a complete
intersection.  In this way the complex structure moduli of $\hat{Z}_3$
naturally form a subsector of the complex structure moduli of $\hat X_4$.  In
summary, the general idea of this discussion is to reformulate and slightly
extend the heterotic/F-theory duality map schematically as:
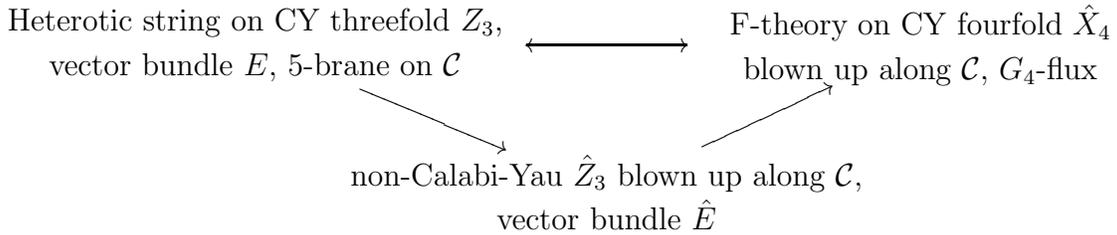
\begin{equation*}
	\xymatrix @C=-1in {
	\parbox{7cm}{\centering Heterotic string on CY threefold $Z_3$,\\ vector bundle $E$, 5-brane on $\mathcal C$} \ar[rd]\ar@{<->}[rr] & & \parbox{6cm}{\centering F-theory on CY fourfold $\hat X_4$\\ blown up along $\mathcal C$, $G_4$-flux}\\
	& \parbox{7cm}{\centering non-Calabi-Yau $\hat Z_3$ blown up along $\mathcal C$,\\ vector bundle $\hat E$} \ar[ru]&
	}
\end{equation*}
where the horizontal arrow indicates the action of heterotic/F-theory duality.

Following this general strategy the paper is organized as follows. In section
\ref{sec:heterotic} we first recall the connection between small instantons and
heterotic five-branes. This allows us to introduce the respective heterotic
superpotentials. Moreover, we discuss the general blow-up procedure of the
heterotic Calabi-Yau threefold, and comment on the representation and
properties of the holomorphic three-form on the blown-up geometry $\hat Z_3$.
In section~\ref{F-theory_sec}, we first review the heterotic/F-theory duality,
highlighting the map of five-branes into a Calabi-Yau fourfold geometry.  We
then discuss the F-theory flux superpotential and describe how it is matched
with its heterotic counterpart. In the last section we study two classes of
examples. Firstly, we discuss the geometrical construction of the heterotic
blow-up threefold and its associated Calabi-Yau fourfold constructed as
complete intersections. Secondly, we investigate an example for which we
explicitly compute the superpotential and confirm the map between five-brane
deformations and fourfold complex structure moduli. 

\newpage

\section{Heterotic Five-Branes and Superpotentials}\label{sec:heterotic}
 
In this section we review the construction of $\cN=1$ vacua by compactification
of the heterotic string on a Calabi-Yau threefold $Z_3$ with vector bundle $E$
and a number of space-time filling five-branes.  We discuss the relation of the
bundle moduli and five-brane deformations via a small instanton transition in
section \ref{hettransition}.  The heterotic superpotential for these moduli
fields will be introduced in section \ref{het_superpot}. It will be argued that
the motion of the five-brane inside $Z_3$ is constraint by a superpotential
given by the integral of the holomorphic three-form $\Omega$ over a chain
ending on the five-brane. In section \ref{heterotic_blowup} we discuss the general 
blow-up procedure and some properties of the resulting geometry.

\subsection{Transition between Heterotic Vector Bundles and Five-Branes}
\label{hettransition}

Let us begin by reviewing the heterotic compactifications.  Besides the choice
of a Calabi-Yau threefold $Z_3$, a consistent heterotic vacuum requires a
choice of stable holomorphic vector-bundles $E=E_1\oplus E_2$ over $Z_3$ which
determine the gauge group preserved in the perturbative $E_8 \times E_8$ of the
heterotic theory. In general we can additionally have five-branes wrapping
holomorphic curves $\cC$ in the threefold $Z_3$. This setup is further
constrained by the general heterotic anomaly cancellation condition
\begin{equation}
	\lambda(E_1)+\lambda(E_2)+\left[\cC \right]=c_2(Z_3)\ ,
\label{eq:anomaly}
\end{equation}
where $\lambda(E)$ is the fundamental characteristic class of the vector bundle
E, which, for example, is $c_2(E)$ for $SU(N)$ bundles and $c_2(E)/60$ for
$E_8$ bundles.  This condition dictates consistent choices of the cohomology
class $\left[\cC \right]$ of the curve $\cC$ in the presence of non-trivial
vector bundles to match the curvature of the threefold $Z_3$ as measured by the
second Chern class $c_2(Z_3)$.  In particular, it implies that $\cC$ corresponds to an effective class in $H_2(Z_3,\mathds{Z})$ \cite{Donagi:1999gc}.

The analysis of the moduli space of the heterotic string on $Z_3$ requires the
study of three a priori  very different pieces.  Firstly, we have the geometric
moduli spaces of the threefold $Z_3$ consisting of the complex structure as
well as K\"ahler moduli space.  Secondly, there are the moduli of the
bundles $E_1$ and $E_2$ which parameterize different gauge-field backgrounds on
$Z_3$. Finally, if the five-brane is wrapped on a non-rigid curves $\cC$, the
deformations of $\cC$ within $Z_3$ of the various five-branes have to be taken
into account.  The entire moduli space is in general very complicated and
difficult to analyze.  This problem, however, becomes more tractable if one
focuses on elliptically fibered Calabi-Yau threefolds $Z_3$. It was shown in
ref.~\cite{Friedman:1997yq} that there exist elegant constructions of the
vector bundle $E$ on these threefolds. Moreover, the moduli space of
five-branes on elliptically fibered $Z_3$ has been discussed in great detail in
ref.~\cite{Donagi:1999jp}.  In general, it admits several different branches
corresponding to the number and type of five-branes present. However, there are
distinguished points in the moduli space corresponding to enhanced gauge
symmetry \cite{Witten:1995gx,Ganor:1996mu} of the heterotic string that allow
for a clear physical interpretation and that we now discuss in more detail. It
will turn out that at these points an interesting transition is possible where a
five-brane completely dissolves into a finite size instanton of the bundle $E$
and vice versa.

Let us start with a threefold $Z_3$ with $c_2(Z_3)\neq 0$ and no five-branes.
Thus, the anomaly condition \eqref{eq:anomaly} forces us to turn on a
background bundle $E$ with non-trivial second Chern class $c_2(E)$ in order to
cancel $c_2(Z_3)$. Then the bundle is topologically non-trivial and carries
bundle instantons characterized by the topological second Chern number
\cite{Witten:1985bz}
\begin{equation}
 	\left[c_2\right]=-\int_{Z_3}J\wedge \mathcal F\wedge \mathcal F\,,
\end{equation}
where $J$ denotes the K\"ahler form on $Z_3$ and $\mathcal F$ the field strength of the background bundle. The heterotic gauge group $G_1$ in
four dimensions is generically broken and given by the commutant of the
holonomy group of the bundle $E$ in $E_8$. Varying the moduli of $E_1$ one can
ask whether it is possible to restore parts or all of the broken gauge symmetry
by flattening out the bundle as much as possible~\cite{Aspinwall:1996mn}.  To
show how this can be achieved, one first decomposes $c_2(E)$ into its components
each of which being dual to an irreducible curve $\cC_i$ in $Z_3$. Since the
invariant $\left[c_2\right]$ has to be kept fixed, the best we can do is to
consecutively split off the components of $c_2(E)$ and to localize the
curvature of $E$ on the corresponding curves $\cC_i$. This should be contrasted
with the generic situation, where the curvature is smeared out all over $Z_3$.
In the localization limit the holonomy of the bundle around each individual
curve $\cC_i$ becomes trivial and the gauge group $G$ enhances accordingly.
Having reached this so-called small instanton configuration at the boundary of
the moduli space of the bundle, the dynamics of (this part of) the gauge bundle
can be completely described by a five-brane on $\cC_i$ \cite{Witten:1995gx}.

Small instanton configurations thus allow for transitions between
branches of the moduli space with different numbers of five-branes, that
consequently map bundle moduli to five-brane moduli and vice versa
\cite{Buchbinder:2002ji}. This is precisely what we need for our later F-theory
analysis.  Note that this transition is completely consistent with
\eqref{eq:anomaly} since we have just shifted irreducible components between the two summands $c_2(E)$
and $\left[\cC \right]$. Thus, we are in the following allowed to think about
the small instanton configuration as the presence of a five-brane. In
particular, doing this transition for all components of $c_2(E)$ the full
perturbative heterotic gauge group $E_8\times E_8$ can be restored.  Turning
this argument around, a heterotic string with full $E_8\times E_8$ gauge
symmetry on a threefold $Z_3$ with non-trivial $c_2(Z_3)$ has to contain
five-branes to cancel the anomaly according to \eqref{eq:anomaly}.  In our concrete example
of section \ref{OurExamples} we will precisely encounter this situation guiding
us to the interpretation of the F-theory flux superpotential in terms of a
superpotential for a particular class of five-branes.

To precisely specify the five-branes we will consider later, we note that on an
elliptically fibered Calabi-Yau threefold the five-brane class
$\left[\cC\right]$ can be decomposed as
\begin{equation}
 	\cC=n_f F+\cC_B\ ,
\end{equation}
where $\cC_B$ denotes a curve in the base $B_2$ of the elliptic fibration, $F$
denotes the elliptic fiber, and $n_f$ is a positive integer.  This is a split
into five-branes vertical to the projection $\pi:\,Z_3\rightarrow B_2$, where
the integer $n_f$ counts the number of five-branes wrapping the elliptic fiber,
and into horizontal five-branes on $\cC_B$ in the base $B_2$.  Both cases are
covered by \eqref{eq:anomaly}, but lead to different effects in the F-theory
dual theory. Vertical five-branes correspond to spacetime filling three-branes
at a point in the base $B_3$ of the F-theory fourfold $X_4$
\cite{Andreas:1997ce,Friedman:1997yq}. Conversely, horizontal five-branes on
the curve $\cC_B$ map completely to the geometry of the F-theory side. They map to
seven-branes supported on a divisor in the fourfold base $B_3$ which
projects onto the curve $\cC$ in $B_2$ \cite{Morrison:1996na, Bershadsky:1996nh,
Bershadsky:1997zs} that has to be blown-up in $B_3$ into a divisor
$D$ \cite{Berglund:1998ej,Rajesh:1998ik,Diaconescu:1999it}. Of course, there can
be mixed types of five-branes as well.  It will be precisely the horizontal
five-branes corresponding to blow-ups into exceptional divisors $D$ for which
our analysis and calculation of the superpotential will be performed.

\subsection{The Heterotic Superpotential}
\label{het_superpot}

The small instanton transition implies a transition between bundle and
five-brane moduli~\cite{Buchbinder:2002ji}. Since both types of moduli are
generally obstructed by a superpotential also the superpotentials for bundle
and five-brane have to be connected by the transition.  As was argued in
\cite{Witten:1997ep} in the context of M-theory on a Calabi-Yau threefold, a
spacetime-filling M5-brane supported on a curve $\cC$ in general induces a
superpotential 
\begin{equation}
\label{eq:chain}
 	W_{\text{M5}}=\int_{\Gamma}\Omega\,,
\end{equation}
where $\Gamma$ denotes a three-chain bounded by $\cC$ and an unimportant
reference curve $\cC_0$ in the homology class of $\cC$. It depends on both the
moduli of the five-brane on $\cC$ as well as the complex structure moduli of
$Z_3$ in the holomorphic three-form $\Omega$.  On the other hand, the
perturbative superpotential for the heterotic bundle moduli is given by the
holomorphic Chern-Simons functional \cite{Witten:1985bz}
\begin{equation}
\label{eq:hCS}
 	W_{\text{CS}}=\int_{Z_3}\Omega\wedge(A\bar{\partial}A+\frac23A\wedge A\wedge A)\,,
\end{equation}
where $A$ denotes the gauge connection that depends on the bundle moduli. The
dependence on the complex structure moduli of $Z_3$ is implicit through
$\Omega$.

To see how the two superpotentials \eqref{eq:chain} and \eqref{eq:hCS} are
mapped onto each other in the transition, let us assume a single instanton
solution $\mathcal{F}$ with $\mathcal{F}\wedge \mathcal{F}$ dual to an
irreducible curve $\cC$. Displaying the explicit moduli dependence of the
configuration $\mathcal{F}$ \cite{Tong:2005un},  in the small instanton limit
$\mathcal{F}\wedge \mathcal{F}$ reduces to the delta function $\delta_{\cC_i}$
of four real scalar parameters. They describe the position moduli of the instanton normal to the curve
in the class $\left[\cC_i\right]$ on which it is localized. Inserting the gauge configuration $\mathcal{F}$
into $W_{\text{CS}}$, the holomorphic Chern-Simons functional is effectively
dimensionally reduced to the curve $\cC$ \cite{Aganagic:2000gs}. In the
vicinity of $\cC$ we may write the holomorphic three-form as $\Omega=d\omega$
which we insert into \eqref{eq:hCS} in the background
$\mathcal{F}\wedge\mathcal{F}$ to obtain
\begin{equation}
 	W_{\rm{CS}}=\int_{\cC}\omega\,
\end{equation}
after a partial integration. Adding a constant given by the integral of
$\omega$ over the reference curve $\cC_0$ this precisely matches the chain
integral \eqref{eq:chain}.  Applying the above discussion, we can think about
the M5-brane moduli in $W_{\rm{M5}}$ as the bundle moduli describing the
position of the instanton configuration $\mathcal{F}$, that in the small
instanton limit precisely map to sections $H^0(\cC_i,N_{Z_3}\cC_i)$ of the
normal bundle to $\cC_i$. 

We will verify this matching of moduli explicitly from the perspective of the
F-theory dual setup later on. There we will on the one hand identify some of
the fourfold complex structure moduli with the heterotic bundle moduli, on the
other hand, however, show that part of the F-theory flux superpotential
depending on the same complex structure moduli really calculates the
superpotential of a five-brane on a curve. This way, employing
heterotic/F-theory duality, we show in the case of an example the equivalence
of the small instanton/five-brane picture. 

To complete the discussion of perturbative heterotic superpotentials, let us
also comment on the flux superpotential due to bulk fluxes. In general, the
heterotic $B$-field can have a non-trivial background field strength $H^{\rm
flux}_3$ that has to be in $H^3(Z_3,\mathds{Z})$ due to the flux quantization
condition. The induced superpotential will be intimately linked to
\eqref{eq:chain} and \eqref{eq:hCS} due to the anomaly cancellation condition 
\beq \label{dH3}
   dH_3 = \tr (\cR \wedge \cR) - \tr (\cF \wedge \cF) - \sum_i \delta_{\cC_i}\ ,
\eeq 
which yields, with an appropriate definition of the traces, 
the condition \eqref{eq:anomaly} if one restricts to cohomology classes.  The
superpotential in terms of this $H_3$ reads \cite{Gukov:1999gr, Behrndt:2000zh}
\begin{equation}
\label{hetallpots}
 	W_{\rm het}=\int_{Z_3}\Omega\wedge H_3= W_{\rm flux} + W_{\rm CS} + W_{\rm M5}\ ,
\end{equation}
where the different terms can be associated to the various contributions in $H_3$
in \eqref{dH3}.
In order to discuss the flux part, we expand $H^{\rm flux}_3=N^i\alpha_i-M_i\beta^i$ in the integral basis $\alpha_i$, $\beta^i$ of $H^3(Z_3,\mathds{Z})$ with integer flux numbers $N^i$, $M_i$.
Then one can write the flux superpotential as
\begin{equation}
\label{eq:fluxpot}
 	W_{\rm{flux}}=\int_{Z_3}\Omega \wedge H_3^{\rm flux}  = M_iX^i-N^iF_i\,,
\end{equation}
where we introduced the period expansion $\Omega=X^i\alpha_i-F_i\beta^i$.  In
general, the periods $(X^i,F_i)$ admit a complicated dependence on the complex
structure deformations of $Z_3$.  It is the great success of algebraic geometry
that this superpotential can be calculated explicitly for a wide range of
examples, see~\cite{Polchinski:1995sm} and \cite{Douglas:2006es,Denef:2008wq} 
for reviews.  This is due to the fact that the periods $X^i$,
$F_i$ obey differential equations, the so-called Picard-Fuchs
equations\footnote{See for example \cite{Hosono:1993qy} for a review.}, that
can be solved explicitly and thus allow to determine the complete moduli
dependence of $W_{\rm{flux}}$.  To end our discussion of the flux
superpotential, let us stress that strictly speaking there is a back-reaction
of $H_3^{\rm flux}$ which renders $Z_3$ to be non-K\"ahler
\cite{Strominger:1986uh}.  Since our main focus will be on the five-brane
superpotential, we will not be concerned with this back-reaction in the
following.

\subsection{The Blow-Up of the Heterotic Calabi-Yau Threefold}
\label{heterotic_blowup}

The form of the superpotential $W_{\text{M5}}$ of \eqref{eq:chain} is rather
universal.  It occurs, for example, also for D5-branes on curves in Type IIB
orientifold compactifications.\footnote{Similar expressions arise for higher
dimensional branes with world-volume flux inducing D5-charge supported on the
same curves.} In the following we will apply the blow-up procedure suggested in
ref.~\cite{Grimm:2008dq} for the study of the chain integral for D5-branes to
the heterotic setup.  The idea is to find a purely geometric description that
puts the dynamics of the five-brane and the geometry of $Z_3$ on an equal
footing.  To achieve this, we blow up the curve $\mathcal{C}$ into a rigid
divisor $D$ in a non-Calabi-Yau threefold $\hat Z_3$. This embeds  the
deformation modes of $\mathcal{C}$ in $Z_3$ as well as the complex structure
deformations of $Z_3$ into the deformation problem of only complex structures
of $\hat Z_3$.  We will see explicitly later that this alternative view on the
heterotic string with five-branes allows for a direct geometric interpretation
of the fate of the five-brane dynamics in heterotic/F-theory duality.  Here we
provide the geometrical tools to describe the blow-up of $Z_3$ along a curve
$\mathcal{C}$ which we will later use in the construction of explicit examples
in section \ref{OurExamples}.

For concreteness, let us consider a Calabi-Yau threefold $Z_3$ described as the
hypersurface $\{P = 0\}$ in a projective or toric ambient space $V_4$. Consider then
a curve $\mathcal{C}$ specified by two additional constraints $\{h_1 = h_2 = 0\}$
in the ambient space intersecting transversally $Z_3$. For the example of horizontal 
curves $\cC$ supporting horizontal five-branes in an elliptic $Z_3$, the constraints 
take the form
\begin{equation}
	h_1 \equiv\tilde{z}=0\ ,\quad h_2\equiv g_5=0\ , 
\end{equation}	
where $\{\tilde{z}=0\}$ restricts to the base $B_2$ and $g_5$ specifies $\cC$ within
$B_2$. In general, the constraints $h_1$, $h_2$
describe divisors in the ambient space that descend to divisors\footnote{The
Lefshetz-Hyperplane theorem tells us that indeed any divisor and line bundle in
$Z_3$ is induced from the ambient space \cite{Griffiths}.} in $Z_3$ as well
upon intersecting with $\{P=0\}$, called $D_1$ and $D_2$.  Locally, $(h_1,h_2)$ can
be considered as normal coordinates to the curve $\cC$ in $Z_3$. Thus, the
normal bundle $N_{Z_3}\mathcal{C}$ of the curve takes the form
$N_{Z_3}C=\mathcal{O}_{Z_3}(D_1)\oplus\mathcal{O}_{Z_3}(D_2)$, where
$\mathcal{O}_{Z_3}(D_i)$ denotes the line bundle of $D_i$ as read of from the
scalings of the section $h_i$.  As the divisors $D_i$, also their line bundles
$\mathcal{O}_{Z_3}(D_i)$ are induced from the bundles $\mathcal{O}(D_i)$ on the
ambient space $V_4$.

To describe the blown-up threefold $\hat Z_3$, we introduce the total space of
the projective bundle $\mathds{P}(\mathcal{O}(D_1)\oplus\mathcal{O}(D_2))$.
This total space describes a $\mathds{P}^1$-fibration over the ambient space
$V_4$ on which we introduce the $\P^1$-coordinates $(l_1,l_2)\sim
\lambda(l_1,l_2)$.  Then, the blow-up $\hat{Z}_3$ is given by the complete
intersection \cite{Griffiths}
\begin{equation}
        P= 0\ ,\qquad 	Q= l_1h_2 - l_2h_1 = 0 \, ,
\label{eq:blowup}
\end{equation}
in the projective bundle.  This is easily checked to describe $\hat{Z}_3$. The
first constraint depending only on the coordinates of the base $V_4$ of the
projective bundle restricts to the threefold $Z_3$. The second constraint then
fibers the $\mathds{P}^1$ non-trivially over $Z_3$ to describe the blow-up
along $\mathcal{C}$. Away from $h_1\neq 0$ or $h_2\neq0$ we can solve
\eqref{eq:blowup} for $l_1$ or $l_2$ respectively.  Thus, \eqref{eq:blowup}
describes a point in the $\mathds{P}^1$-fiber for every point in $Z_3$ away
from the curve. However, if $h_1=h_2=0$ the coordinates $(l_1,l_2)$ are
unconstrained and parameterize the full $\mathds{P}^1$, which is fibered over
$\mathcal{C}$ as its normal bundle $N_{Z_3}\mathcal{C}$.  Thus, we have
replaced the curve by the exceptional divisor $D$ that is given by the
projectivization of its normal bundle in $Z_3$, i.e.~the ruled surface
$D=\P(N_{Z_3}\mathcal C)$ over $\mathcal{C}$. We denote the blow-down map by 
\begin{equation}
	\xymatrix{\pi:\hat Z_3\ar[r] & Z_3}\,.
\end{equation}

Having described the construction of the blow-up, one can also determine
details on the cohomology of $D$ and $\hat Z_3$ \cite{Griffiths}.  For a single
smooth curve $\cC$ the non-vanishing Hodge numbers of $D$ are determined to be
\beq
       h^{0,0}= h^{2,2}=1\ , \qquad h^{1,0}=g\ ,\qquad h^{1,1}=2
       \label{eqn:hodge-exceptional-divisor}
\eeq
as usual for a ruled surface $D$ over a genus $g$ curve $\mathcal C$. One
element, which we denote by $\eta|_D$, of $H^{1,1}(D)$ is induced from the
ambient space $\hat Z_3$ and given by $\eta=c_1(N_{\hat Z_3}D)$. The second
element spanning $H^{1,1}(D)$ is given by the Poincar\'e dual $[\cC]_D$ of the
curve $\cC$ in $D$, $[\cC]_D =c_1(N_D \cC)$. It is related to the first Chern
class $c_1(\mathcal{C})$ and thus to the genus as
\begin{equation}
	c_1(N_D\mathcal{C})=-c_1(\cC)-2\eta\,,
\end{equation}
by using the adjunction formula in $\hat{Z}_3$. Note that as a blow-up divisor 
$D$ is rigid in $\hat Z_3$. 
The first and second Chern class of $\hat{Z}_3$ are affected by the blow-up as
\begin{eqnarray}
	c_1(\hat Z_3) &=&\pi^*(c_1(Z_3))-c_1(N_{\hat Z_3}D)\ ,\\
         c_2 (\hat Z_3) &=& \pi^*(c_2(Z_3) + [\cC]) - \pi^*(c_1(Z_3)) D\ .
\label{eqn:c1-ztilde}
\end{eqnarray}
Clearly, if $Z_3$ is a Calabi-Yau manifold one can use $c_1(Z_3)=0$ to find
\beq
   c_1 (\hat Z_3) = - \eta \ , \qquad   c_2 (\hat Z_3) = \pi^*(c_2(Z_3) + [\cC]) \ ,
\eeq
in particular that $\hat{Z}_3$ is no more Calabi-Yau.

It was argued in \cite{Grimm:2008dq} that the complex structure moduli space of
$\hat{Z}_3$ contains the complex structure moduli of $Z_3$ as well as the
deformation of $\mathcal{C}$ within $Z_3$.  The basic reason for this is
roughly that the complex structure deformations of the rigid divisor $D$
contain the deformation moduli of the curve $\cC$ and thus embed them into the
complex structure of $\hat{Z}_3$. This way the deformations of the pair
$(Z_3,\cC)$ form a subsector of the geometrical deformations of $\hat{Z}_3$.
This allows for the study of the combined superpotential of five-brane
\eqref{eq:chain} and flux \eqref{eq:fluxpot} as well.  First we use the formal
unification of the two superpotentials in terms of the relative homology group
$H_3(Z_3,\mathcal{C},\mathds{Z})$ consisting of three-cycles
$H_3(Z_3,\mathds{Z})$ and three-chains $\Gamma^{\cC}$ ending on the curve
$\cC$.  Then the superpotential can be written as \cite{Lerche:2002ck}
\begin{equation}
      W_{\rm flux} + W_{\rm M5} =\sum_i \tilde N^i \int_{\Gamma^i_\cC} \Omega \,
\label{eq:relsup}
\end{equation}
with respect to an integral basis $\Gamma^i_\cC$ of the relative group
$H_3(Z_3,\cC,\mathds{Z})$.  Here the integers $\tilde N^i$ correspond to the
three-form flux quanta $(M_i,N^i)$ in \eqref{eq:fluxpot} and the five-brane
windings.  In particular $\Omega$ has to be interpreted as a relative form.

It has been argued in ref.~\cite{Grimm:2008dq} that in the blow-up
$\pi:\hat{Z}_3\rightarrow Z_3$ the superpotential \eqref{eq:relsup} is lifted
to $\hat{Z}_3$ as follows.  First we have to replace $\Omega$ by its equivalent
on $\hat{Z}_3$, the pullback form
\begin{equation}
	\hat{\Omega}=\pi^*(\Omega)\,,\quad \hat{\Omega}|_D=0
\label{eq:proptr}
\end{equation}
that can be shown to vanish on $D$, see \cite{Grimm:2008dq} for details and
references.  Consequently we can write the heterotic superpotentials as
\begin{equation}
 	W_{\rm flux}+W_{\rm M5}=\int_{\hat{Z}_3} H_3\wedge
\hat{\Omega}=\int_{\hat{Z}_3-D}H_3\wedge\hat{\Omega}=\int_{\Gamma_{H_3}}\hat{\Omega}
\end{equation}
such that it only depends on the topology of the open manifold
$Z_3-\cC=\hat{Z}_3-D$. Here, we naturally obtain $\Gamma_{H_3}$ as the
Poincar\'e dual of the flux $H_3$ in the group $H_3(\hat{Z}_3-D,\mathds{Z})$.

These replacements can also be understood in the language of relative
(co)homology. On the one hand we can treat $\hat{\Omega}$ as a relative form
exploiting the fact that any element in the relative group
$H^{3}(\hat{Z}_3,D,\mathds{Z})$ can be represented by a form vanishing on $D$.
On the other hand the element $\Gamma_{H_3}$ maps to the relative homology
since Lefshetz and Poincar\'{e} duality relate the de Rham homology of the open
manifold to the relative homology as
\begin{equation}
	H_{3}(\hat Z_3 -D,\mathds{Z})=H_3(\hat{Z}_3,D,\mathds{Z})
\label{eq:Lefshetz}
\end{equation}
This identification of (co-)homology groups gets completed by the
equivalence $H_3(Z_3,\cC,\mathds{Z})=H_3(\hat{Z}_3,D,\mathds{Z})$ telling us
that we have consistently replaced all relevant topological quantities on $Z_3$
by those on the blow-up $\hat{Z}_3$.  Finally, we expand the element
$\Gamma_{H_3}$ in a basis $\Gamma^i_D$ of
$H_3(\hat{Z}_3-D,\mathds{Z})=H_3(\hat{Z}_3,D,\mathds{Z})$ to obtain an
expansion of the superpotential by relative periods of $\hat{\Omega}$ as
\begin{equation}
	 W_{\rm flux}+ W_{\rm M5}=\sum_i \tilde {N}^i \int_{\Gamma^i_D}
\hat{\Omega} = \sum_i  \tilde {N}^i \int_{\hat Z_3} \hat \Omega \wedge
  \gamma_i^{D}\ .
\label{eq:liftW}
\end{equation}
Here $ \gamma_i^{D}$ are the Poincar\'e duals in
$H^{3}(\hat{Z}_3,D,\mathds{Z})$.

Similar to the Calabi-Yau threefold case where every element in
$H^3(Z_3,\mathds{Z})$ can be obtained upon differentiating $\Omega$ with
respect to the complex structure, it is possible to obtain a basis of
$H^{3}(\hat{Z}_3,D,\mathds{Z})$ the same way. More precisely we can write the
basis elements $\gamma_i^{D}$ as differentials of $\hat \Omega$ evaluated at
the large complex structure point,
\beq
   \gamma_i^{D} = \mathcal R_i \hat \Omega|_{\underline z=0} \ .
\eeq
The operators $\cR_i$ are polynomials in the differentials $\theta_a =
z_a\frac{d}{dz_a}$.  Such a representation can be made explicit by noting that
$\hat \Omega$ can be written as a residue integral \cite{Griffiths1}
\beq
\label{ResZhat}
\hat \Omega =
\int_{\epsilon_1}\int_{\epsilon_2} \frac{\Delta}{P Q}\ ,
\eeq
where $P,Q$ are the two constraints \eqref{eq:blowup} which define $\hat Z_3$.
The form $\Delta$ denotes a top-form on the five-dimensional ambient space
$\mathds{P}(\mathcal{O}(D_1)\oplus\mathcal{O}(D_2))$ that is invariant under
its torus actions and the $\epsilon_i$ are loops around $\{P=0\}$, $\{Q=0\}$. For the
type of ambient space we consider, the measure $\Delta$ takes the schematic
form \cite{Witten:1996bn}
\begin{equation}
 	\Delta=\Delta_V\wedge (l_1dl_2-l_2dl_1)\,,
\end{equation}
where $\Delta_V$ denotes the invariant form on the toric base $V_4$ and
$(l_1,l_2)$ the coordinates of the $\mathds{P}^1$-fiber.  This makes it
possible to study some of the afore-mentioned properties of $\hat{\Omega}$
explicitly.

The crucial achievement of the blow-up to $\hat{Z}_3$ is the fact that all
moduli dependence of the superpotential is now contained in the complex
structure dependence of $\hat{\Omega}$. Thus it is possible, analogous the
Calabi-Yau case, to derive Picard-Fuchs type differential equations for
$\hat{\Omega}$ by studying its complex structure dependence explicitly. Upon
the algebraic representation of $\hat{Z}_3$ by the complete intersection
\eqref{eq:blowup} it is now possible to find an explicit residue representation
of $\hat{\Omega}$ such that Griffiths-Dwork reduction can be used to derive the
desired differential equations for $\hat{\Omega}$, among whose solutions we
find the superpotential $W$.

So far the discussion of the blow-up procedure and the determination of the
brane and flux superpotential was entirely in the heterotic theory $Z_3$.
However, we will shed more light on the connection between the brane geometry
of $(Z_3,\cC)$ and the classical complex geometry of the blow-up $\hat{Z}_3$ in
the context of heterotic/F-theory duality.  More precisely, we argue that the
five-brane superpotential is mapped to a flux superpotential for F-theory
compactified on a dual Calabi-Yau fourfold $\hat X_4$.  Starting with $\hat
Z_3$, the fourfold $\hat X_4$ can be represented as a complete intersection
generalizing \eqref{eq:blowup}. However, in contrast to $\hat Z_3$ the fourfold
$\hat X_4$ can also be represented as a hypersurface. This fact allows us to
directly compute the flux superpotential.  Such a computation has been
performed in ref.~\cite{Grimm:2009ef} for a set of examples, and confirmed that
the five-brane superpotential is naturally contained in the F-theory flux
superpotential.  In the next section we will discuss this duality in detail and
outline the construction of $\hat X_4$ and the  F-theory flux $G_4$.

\section{F-Theory Blow-Ups and the Superpotential}
\label{F-theory_sec}

Here we turn to the discussion of F-theory compactifications on elliptic
Calabi-Yau fourfolds $X_4$ yielding $\mathcal{N}=1$ effective theories in four
dimensions. We will discuss the basic geometric ingredients encoding the
seven-brane content as well as the three-brane tadpole in its most general form
including $G_4$-flux in section \ref{F_basics}. There, we will readily restrict to 
F-theory fourfolds $X_4$ with a heterotic dual on an elliptic threefold $Z_3$. The F-theory dual to an
$E_8\times E_8$ heterotic string with small instantons/five-branes is discussed in section \ref{F_blowup}
requiring a blow-up in the F-theory base $B_3$ along curves $\mathcal{C}$ in
$B_2$. We will argue that the five-brane moduli and superpotential are mapped
to complex structure moduli of $X_4$ and the flux superpotential. Finally in section \ref{DualSupP}, we
will construct the appropriate $G_4$-flux inducing the flux superpotential dual
to the heterotic brane superpotential.

\subsection{F-Theory and Heterotic/F-Theory Duality}
\label{F_basics}

We prepare for our further discussion by briefly reviewing the necessary aspects of F-theory and heterotic/F-theory duality.

An F-theory compactification to four dimensions is in general defined by an elliptically fibered Calabi-Yau fourfold $X_4$ with a section. This section can be used to express the fourfold $X_4$ as an analytic hypersurface in the projective bundle $\mathds{P}(\mathcal{O}_{B_3}\oplus \mathcal{L}^2\oplus \mathcal{L}^3)$ with coordinates $(z,x,y)$ for which the constraint equation can be brought to the Weierstrass form
\begin{equation}
\label{eq:wsf}
 	y^2=x^3+fxz^4+gz^6\,.
\end{equation}
The Calabi-Yau condition on $X_4$ implies $\mathcal{L}=K^{-1}_{B_3}$ and $f$, $g$ have to be sections of $\mathcal{L}^4$ and $\mathcal{L}^6$ for the constraint \eqref{eq:wsf} to transform as a section of $\mathcal{L}^6$.
F-theory defined on $X_4$ automatically takes care of a consistent inclusion of spacetime-filling seven-branes. These are supported on the in general reducible divisors $\mathbf{\Delta}$ in the base $B_3$ determined by the degeneration loci of \eqref{eq:wsf} given by the discriminant
\begin{equation}
 	\mathbf{\Delta}=\{\Delta=27g^2+4f^3=0\}\,.
\end{equation}
The degeneration type of the fibration specified by the order of vanishing of $f$, $g$ and $\Delta$ along the irreducible components $\mathbf{\Delta}_i$ of the discriminant have an ADE--type classification that physically specifies the four-dimensional gauge group $G$ \cite{Bershadsky:1996nh}. 

There are further building blocks necessary to specify a consistent F-theory setup. This is due to the fact that a four-dimensional compactification generically has a three-brane tadpole of the form \cite{Vafa:1995fj,Becker:1996gj,Sethi:1996es}
\begin{equation}
\frac{\chi(X_4)}{24}=n_3+\frac{1}{2}\int_{X_4}G_4\wedge G_4\,.
\label{eq:3tadpole}
\end{equation}
In the case that the Euler characteristic $\chi(X_4)$ of $X_4$ is non-zero a given number $n_3$ of spacetime-filling three-branes on points in $B_3$ and a specific amount of quantized four-form flux $G_4$ have to be added in order to fulfill \eqref{eq:3tadpole}.

For a generic setup with three-branes and flux, the four-dimensional gauge symmetry as determined by the seven-branes is not affected. However, if the three-brane happens to collide with a seven-brane, it can dissolve, by the same transition as discussed in section \ref{hettransition}, into a finite-size instanton on the seven-brane worldvolume that breaks the four-dimensional gauge group $G$. During this transition the number $n_3$ of three-branes jumps and a flux $G_4$ is generated describing the gauge instanton on the seven-brane worldvolume \cite{Denef:2008wq}. 
In particular, in case of a heterotic dual theory the three-branes on the F-theory side precisely correspond to vertical five-branes on the heterotic threefold \cite{Friedman:1997yq}. Thus, under duality the three-brane/instanton transition is precisely the F-theory dual of the transition of a vertical five-brane into  
a finite size instanton breaking the gauge group on the heterotic side accordingly.
However, we will not encounter this any further since we restrict our discussion to the case that the gauge bundle on those seven branes dual to the perturbative heterotic gauge group is trivial and no three-branes sit on top of their worldvolumes. 

Let us now come to a more systematic discussion of heterotic/F-theory duality.
The fundamental duality that underlies it in any dimensions is the eight-dimensional equivalence of the heterotic string compactified on $T^2$ and F-theory on elliptic K3 \cite{Vafa:1996xn}. The eight-dimensional gauge symmetry $G$ is determined in the heterotic string as the commutant of an $E_8\times E_8$-bundle on $T^2$ with structure group $H$. This precisely matches the singularity type $G$ of the elliptic fibration of K3 in the F-theory formulation.
Using the adiabatic argument \cite{Vafa:1995gm} it is possible to consider a family of dual eight-dimensional theories parameterized by a base manifold $B_n$ to obtain dualities between the heterotic string and F-theory in lower dimensions. 

This way a four-dimensional heterotic string on the elliptic threefold $Z_3$ is equivalent to F-theory on the elliptic K3-fibered Calabi-Yau fourfold $X_4$. Consequently, the three-dimensional base $B_3$ of the elliptic fibration of $X_4$ has to be ruled over the base $B_2$ of the heterotic threefold $Z_3$, i.e.~$B_3$ is a holomorphic $\mathds{P}^1$-fibration over $B_2$. It turns out that precisely this fibration data of $B_3$ is crucial for the construction of the dual heterotic theory, in particular the stable vector bundle $E$ on $Z_3$ that determines the four-dimensional gauge group $G$.
To analyze this issue in a more refined way it is necessary to use the methods developed in \cite{Friedman:1997yq}, in particular the spectral cover.
However, instead of delving into the technical details, we will focus on the results essential for our further discussion.

The basic strategy of the spectral cover is to obtain the stable holomorphic
bundle $E$ on the elliptic threefold $Z_3$ roughly speaking by fibering the
stable bundles on the fiber torus so that they globally fit into a stable
bundle on the threefold $Z_3$ \cite{Friedman:1997yq}. This way, the topological
data of the bundle $E$ can be determined in terms of the cohomology of the
two-dimensional base $B_2$. For example, for our case of interest, $H=SU(n)$
and $E_8$,\footnote{Strictly speaking, there is no spectral cover description of
$E_8$ bundles. However, upon application of the method of parabolics very
similar results to the $SU(n)$ case can be obtained \cite{Friedman:1997yq}.}
the second Chern class $c_2(E)$ of the bundle $E$ schematically reads
\begin{equation}
	\lambda(E)=\eta\sigma+\pi^*(\omega)\,,
\label{eq:c2E}
\end{equation}  
where $\eta$ and $\omega$ are up to now arbitrary classes in
$H^2(B_2,\mathds{Z})$ and $\sigma=c_1(\mathcal{O(\sigma)})$ is Poincar\'e dual
to the section $\sigma$ of $\pi:Z_3\rightarrow B_2$. The class $\eta$ is essential in the
general construction of the spectral cover. However, its meaning is further
clarified in heterotic/F-theory duality, where it can be constructed from the
base $B_3$ of the dual F-theory.

Consider the heterotic string with an $E_8\times E_8$-bundle on $Z_3$. 
Besides the required singularities of the elliptic fibration of $X_4$ to match the heterotic gauge group $G$ only the part $B_2$ of the F-theory geometry is fixed by duality. The threefold $B_3$ can be freely specified by choosing the $\mathds{P}^1$-fibration over $B_2$ as follows. Fixing a line bundle $L$ over $B_2$ the threefold $B_3$ is described as the total space of the projective bundle $\mathds{P}(\mathcal{O}\oplus L)$. There are two distinguished classes in $H^2(B_3,\mathds{Z})$, namely the $B_2$-independent class of the hyperplane of the $\mathds{P}^1$-fiber denoted by $r=c_1(\mathcal{O}(1))$ and the line bundle $L$ with $c_1(L)=t$. 
Then, the heterotic bundle $E=E_1\times E_2$ is specified by \cite{Friedman:1997yq}
\begin{equation}
	\eta(E_1)=6c_1+t\,,\quad \eta(E_2)=6c_1-t\,,
\label{eq:etaE8}
\end{equation}
meaning that the choice of $\mathds{P}^1$-fibration uniquely determines the $\eta$-classes of the two bundles. In particular, we note that the heterotic anomaly \eqref{eq:anomaly} is trivially fulfilled without the inclusion of any horizontal five-branes.

So far, the above discussion is not the most general setup possible since it does not allow for the presence of horizontal five-branes. It turns out that the F-theory dual to the $E_8\times E_8$ heterotic string has to a be analyzed more thoroughly in order to naturally include horizontal five-branes to the setup.

\subsection{The Five-Brane Dual: Blowing Up in F-Theory}
\label{F_blowup}

In this section we will discuss the F-theory dual of
horizontal five-branes \cite{Rajesh:1998ik,Diaconescu:1999it} as will be
essential for our understanding of the five-brane superpotential.

Thus, we now restrict our considerations completely to F-theory
compactifications with a heterotic dual. Then $B_3$ is the total space of the
projective bundle $\mathds{P}(\mathcal{O}_{B_2}\oplus L)$ where we now assume
$L=\mathcal{O}_{B_2}(-\Gamma)$ for an effective divisor $\Gamma$ in $B_2$.
This fibration $p:B_3\rightarrow B_2$ has two holomorphic sections denoted
$C_0$, $C_\infty$ with
\begin{equation}
 	C_\infty=C_0+p^*\Gamma\,.
 	\label{eq:cinfty}
\end{equation}
Then, the perturbative gauge group $G=G_1\times G_2$, where we denote the group
factors from the first $E_8$ as $G_1$ and from the second $E_8$ as $G_2$, is
realized by seven-branes over $C_0$ and $C_\infty$ with singularity type $G_1$
and $G_2$, respectively \cite{Morrison:1996na,Bershadsky:1997zs}. On the other hand, components
of the discriminant on which $\Delta$ vanishes of order greater than one that
project onto curves $\mathcal{C}_i$ in $B_2$ correspond to heterotic
five-branes on the same curves in $Z_3$
\cite{Morrison:1996na,Bershadsky:1996nh,Bershadsky:1997zs}.
Consequently, the corresponding seven-branes induce a gauge symmetry that is a
non-perturbative effect due to five-branes on the heterotic side.

Since the understanding of horizontal five-branes is the central point of our
discussion let us analyze the consequences of these vertical components of the
discriminant for the F-theory geometry more thoroughly. Guided by our example
of section \ref{Example2}, we will consider the enhanced symmetry point with
$G=E_8\times E_8$ due to small instantons/five-branes such that the heterotic
bundle is trivial. In general, an analysis of the local F-theory geometry near
the five-brane curve $\mathcal{C}$ is possible \cite{Diaconescu:1999it}
applying the method of stable degeneration
\cite{Aspinwall:1997ye,Friedman:1997yq}. However, since the essential point in
the analysis is the trivial heterotic gauge bundle, the results of
\cite{Diaconescu:1999it} carry over to our situation immediately.

As follows in general using \eqref{eq:cinfty} the canonical bundle of the
ruled base $B_3$ reads
\begin{equation}
	K_{B_3}=-2C_0+p^*(K_{B_2}-\Gamma)
	=-C_0-C_\infty+p^*(K_{B_2})\,.
\label{eq:ccB3}
\end{equation}
From this we obtain the classes $F$, $G$ and $\mathbf{\Delta}$ of the divisors
defined by $f$, $g$ and $\Delta$ as sections of $K_{B_3}^{-4}$, $K_{B_3}^{-6}$ and $K_{B_3}^{-12}$, respectively.
To match the heterotic gauge symmetry $G=E_8\times E_8$, there have to be $II^*$
fibers over the divisors $C_0$, $C_\infty$ in $B_3$. Since $II^*$ fibers
require that $f$, $g$ and $\Delta$ vanish to order $4$, $5$ and $10$ over $C_0$
and $C_\infty$, their divisor classes split accordingly with
remaining parts 
\begin{eqnarray}
	F'&=&F-4(C_0+C_\infty)=-4p^*(K_{B_2})\,,\nn\\
	G'&=&G-5(C_0+C_\infty)=C_0+C_\infty-6p^*(K_{B_2})\,,\\
	\mathbf{\Delta}'&=&\mathbf{\Delta}-10(C_0+C_\infty)=2C_0+2C_\infty-12p^*(K_{B_2})\,.\nn
\label{eq:rest}
\end{eqnarray}
This generic splitting implies that the component $\Delta'$ can locally be
described as a quadratic constraint in a local normal coordinate $k$ to $C_0$ or
$C_\infty$, respectively. Thus, $\Delta'$ can be understood locally as a double
cover over $C_0$ respectively $C_\infty$ branching over each irreducible curve
$\mathcal{C}_i$ of $\Delta'\cdot C_0$ and $\Delta'\cdot C_\infty$. In fact, near
one irreducible curve $\cC_i$ intersecting say $C_0$ the splitting \eqref{eq:rest} implies
that the sections $f,g$ take the form 
\beq
\label{eqn:brane-g5}
	f=k^4f'\,,\quad g=k^5(g_5+k g_6)\equiv k^5 g'
\eeq
with $f'$ denoting a section of $KB_3^{-4}$ and $g_5$, $g_6$ sections of $KB_3^{-6}\otimes L$, $KB_3^{-6}$, respectively. 
The discriminant then takes the form $\Delta=k^{10}\Delta'$ where $\Delta'$ is calculated from $f'$ and $g'$.
Thus, the intersection curve is given by $g_5=0$ and the degree of the discriminant $\Delta$ rises by two over $\cC_i$ with
$f'$ and $g'$ vanishing of order zero and one. Precisely the singular curves 
$\mathcal{C}_i$ in $X_4$ that appear in $g$ as above are the locations of the small instantons/horizontal
five-branes in $Z_3$ \cite{Rajesh:1998ik,Diaconescu:1999it} on the heterotic
side. In the fourfold $X_4$ the collision of a $II^*$ and a $I_1$ singularity
over $\mathcal{C}_i$ induces a singularity of $X_4$ exceeding Kodaira's
classification of singularities.  Thus, it requires a blow-up
$\pi:\tilde{B}_3\rightarrow B_3$ in the three-dimensional base of the curves
$\mathcal{C}_i$ into divisors $D_i$. This blow-up can be performed without
violating the Calabi-Yau condition since the shift in the canonical class of
the base, $K_{\tilde{B}_3}=\pi^*K_{B_3}+D_i$, can be absorbed into a
redefinition of the line bundle $\mathcal{L}'=\pi^*\mathcal{L}-D_i$ entering
\eqref{eq:wsf} such that
$K_{X_4}=p^*(K_{B_3}+\mathcal{L})=p^*(K_{\tilde{B}_3}+\mathcal{L}')=0$. 

To describe this blow-up explicitly let us restrict to the local neighborhood 
of one irreducible curve $\mathcal{C}_i$ of the intersection of $\mathbf{\Delta}$ and $C_0$. 
We note  that the curve $\cC_i$ in $B_2$ 
is given by the two constraints
\begin{equation}
	h'_1\equiv k=0\ ,\quad h'_2\equiv g_5=0\ ,
\label{eq:curveB2}
\end{equation}
for $k$ and $g_5$ being sections of the normal bundle $N_{B_3}C_0$ and of $KB_3^{-6}\otimes L$, respectively. 
Then if $X_4$ is given as a hypersurface $P'=0$ we obtain the blow-up as the complete intersection 
\beq \label{4fold_blowup}
  P' = 0 \ , \qquad Q' = l_1 h_2' - l_2 h_1' = 0 \ ,
\eeq 
where, as in \eqref{eq:blowup}, we have introduced coordinates $(l_1,l_2)$ parameterizing the $\mathds{P}^1$-fiber. 

However, at least in a local description, we can introduce a local normal coordinate $t$ to $\cC_i$ in $B_2$ such that 
$g_5=tg_5'$ for a section $g_5'$ which is non-vanishing at $t=0$. Then by choosing a local coordinate $k_1$ of the $\mathds{P}^1$-fiber we can solve the blow-up relation $Q'$ of \eqref{4fold_blowup} to obtain $k=k_1 t$. This coordinate transformation 
can be inserted into the constraint $P'=0$ of $X_4$ to obtain the blown-up fourfold $\hat{X}_4$ as a hypersurface. The $f',g'$ of this hypersurface are given by
\beq
  \label{f'g'}
  f' = k_1^4 f \ , \qquad g' = k_1^5 (g_5 + k_1 t\, g_6 + \ldots)
\eeq  
In particular, calculating the discriminant $\Delta'$ of $\hat{X}_4$ it can be demonstrated that the $I_1$ singularity no longer hits the $II^*$ singularity over $C_0$ \cite{Diaconescu:1999it}.
This way we have one description of $\hat{X}_4$ as the complete intersection \eqref{4fold_blowup} and another as a hypersurface. Both will be of importance for the explicit examples discussed in sections \ref{Example1}, \ref{Example2} 
and in particular section \ref{non-CYblowup}. 

To draw our conclusions of this blow-up, we summarize what we just discussed.
The F-theory counterpart of a heterotic string with full perturbative gauge
group is given by a fourfold with $II^*$ fibers over the sections $C_0$,
$C_\infty$ in $B_3$. The component $\Delta'$ of the discriminant enhances the
degree of $\Delta$ on each intersection curve $\cC_i$ such that a blow-up in
$B_3$ becomes necessary. On the other hand, each blow-up corresponds to a small
instanton in the heterotic bundle
\cite{Morrison:1996na,Aspinwall:1996mn}, that we previously
described in section \ref{hettransition} as a horizontal five-brane on the
curve $\mathcal{C}_i$ in the heterotic threefold $Z_3$ . Indeed, this can be
viewed as a consequence of the observation mentioned above that a vertical
component of the discriminant with degree greater than one corresponds to a
horizontal five-brane \cite{Bershadsky:1997zs} as the degree of $\Delta'$ on
$C_0$ and $C_\infty$ is two.

We finish this discussion by a brief look at the moduli map between F-theory
and its heterotic dual, where we focus on the fate of the five-brane moduli in
the just mentioned blow-up process. The first step in the moduli analysis is to
relate the dimensions of the various moduli spaces in both theories and to
point to possible mismatches where moduli of some ingredients are missing. In
particular, this happens in the presence of heterotic five-branes. Indeed it
was argued in \cite{Rajesh:1998ik} that the relation of the fourfold Hodge
numbers $h^{3,1}(X_4)$ and $h^{1,1}(X_4)$ counting complex structure and
K\"ahler deformations, respectively, to $h^{2,1}(Z_3)$, $h^{1,1}(Z_3)$ and
the bundle moduli and characteristic data has to be modified in the presence of
five-branes. The extra contribution is due to deformation moduli of the curve
$\mathcal{C}_i$ supporting the five-brane counted by
$h^{0}(\mathcal{C}_i,N_{Z_3}\mathcal{C}_i)$ as well as the blow-ups in $B_3$
increasing $h^{1,1}(B_3)$ such that we obtain \cite{Rajesh:1998ik}
\begin{eqnarray}
	h^{3,1}(X_4)&=&h^{2,1}(Z_3)+I(E_1)+I(E_2)+h^{2,1}(X_4)+1+\sum_i h^{0}(\mathcal{C}_i,N_{Z_3}\mathcal{C}_i)\,,\nn\\
		h^{1,1}(X_4)&=&1+h^{1,1}(B_3)+\rm{rk}(G)\,.
\label{eq:modulimap}
\end{eqnarray} 
Here the sum index $i$ runs over all irreducible curves $\mathcal{C}_i$ and we
denote the rank of the four-dimensional gauge group by rk$(G)$. The index $I(E_{1,2})$
counts a topological invariant of the
bundle moduli and is given by \cite{Friedman:1997yq,Andreas:1999zv}
\beq
	I(E_i) = \operatorname{rk}(E_i)+\int_{
	B_2}\left(4(\eta_i\sigma-\lambda_i)+\eta_i
	c_1(B_2)\right)\ .
	\label{eqn:index}
\eeq

The map for $h^{3,1}(X_4)$ reflects the fact that the
four-dimensional gauge symmetry $G$ is on the heterotic side determined by the
gauge bundle $E$ whereas on the F-theory side $G$ is due to the seven-brane
content defined by the discriminant $\Delta$ that is sensitive to a change of
complex structure. For an explicit demonstration of this map exploiting the
techniques of \cite{Berglund:1998ej} we refer to our work \cite{Grimm:2009ef}.

Let us now discuss how \eqref{eq:modulimap} changes during the blow-up procedure. To actually perform the blow-up along the curve $\mathcal{C}_i$ it is necessary to first degenerate the constraint of $X_4$ such that $X_4$ develops the singularity over $\mathcal{C}_i$ described above. This requires a tuning of the coefficients entering the fourfold constraint thus restricting the complex structure of $X_4$ accordingly which means $h^{3,1}(X_4)$ is lowered. Then, we perform the actual blow-up by introducing the new K\"ahler class associated to the complexified volume of the exceptional divisor $D_i$. Thus, we end up with a new fourfold with decreased $h^{3,1}$ and $h^{1,1}(\tilde{B_3})$ increased by one. This is also clear from the general argument \cite{Bershadsky:1997zs} that, enforcing a given gauge group $G$ in four dimensions, the complex structure moduli have to respect the form of $\Delta$ dictated by the singularity type $G$. Since the blow-up which is dual to the heterotic small instanton/five-brane transition enhances the gauge symmetry $G$, the form of the discriminant becomes more restrictive, thus fixing more complex structures. In this picture the blow-down can be understood as switching on moduli decreasing the singularity type of the elliptic fibration.

Similarly, we can understand \eqref{eq:modulimap} from the heterotic side. For each transition between small instanton and five-brane, the bundle loses parts of its moduli since the small instanton is on the boundary of the bundle moduli space. Consequently, the index $I$ reduces accordingly. In the same process, the five-brane in general gains position moduli counted by $h^{0}(\mathcal{C}_i,N_{Z_3}\mathcal{C}_i)$, that have to be added to \eqref{eq:modulimap}.

We close the discussion of moduli by making a more refined and illustrative statement about the heterotic meaning of the K\"ahler modulus of the exceptional divisors $D_i$. To do so we have to consider heterotic M-theory on $Z_3\times S^1/\mathds{Z}_2$. In this picture the instanton/five-brane transition can be understood \cite{Douglas:1996xp} as a spacetime-filling five-brane wrapping $\mathcal{C}_i$ and moving on $S^1/\mathds{Z}_2$ onto the end-of-the-world brane where one perturbative $E_8$ gauge group is located. There, it dissolves into a finite size instanton of the heterotic bundle $E$. With this in mind the distance of the five-brane on the interval $S^1/\mathds{Z}_2$ away from the end-of-world brane precisely maps \cite{Diaconescu:1999it} to the K\"ahler modulus of the divisor $D_i$ resolving $\mathcal{C}_i$ in $B_3$.

\subsection{The F-Theory Flux Superpotential}
\label{Ftheoryfluxpotential}
 
In this section we discuss the F-theory flux superpotential and recall how 
mirror symmetry for Calabi-Yau fourfolds allows to compute its explicit form
\cite{Greene:1993vm,Mayr:1996sh,Klemm:1996ts,Grimm:2009ef}.
Recall, that the F-theory superpotential is induced by four-form flux $G_4$ and 
given by \cite{Gukov:1999ya} 
\begin{equation}
\label{4ffluxsuperpotential} 
W_{G_4}(\underline{t})=\int_{X_4} G_4\wedge \Omega_4(\underline{t})=
N^a\,  \Pi^b(\underline{t})\, \eta_{ab}, \qquad a,b=1,\ldots b^4(X_4)\ ,
\end{equation} 
where $\underline{t}$ collectively denote the $h^{3,1}(X_4)$ complex structure deformations 
of $X_4$. Note that in order to compute $W_{G_4}$ it is necessary to expand 
in a basis $\gamma^a$ of the integral homology group $H_{4}(X_4,\Z)$. 
The $N^a=\int_{\gamma_a} G_4 \in \mathds{Z}/2$ are the flux quantum numbers in this 
basis, while $\Pi^a(\underline{t})=\int_{\gamma^a} 
\Omega_4(\underline{t})$ are the periods of the  holomorphic $(4,0)$-form $\Omega_4$.
The constant intersection matrix 
\beq
   \eta_{ab}=\int_{X_4} \hat \gamma_a \wedge
\hat \gamma_b\ ,\qquad \quad \int_{\gamma_a} \hat \gamma_b=\delta^a_{b}\ ,
\eeq 
is defined for the integral basis $\hat \gamma_a$ of the
cohomology group $H^4(X_4,\mathds{Z})$ which is dual to $\gamma_a$. 
Note that in contrast to $H^3(Z_3,\Z)$ of Calabi-Yau threefolds
the fourth cohomology group of $X_4$ does not carry a symplectic 
structure which necessitates the introduction of $\eta_{ab}$. 
The last expression in
formula (\ref{4ffluxsuperpotential}) is therefore obtained by expanding
$G_4=N^a \hat
\gamma_a$  and $\Omega_4(\underline{t})=\Pi^a(\underline{t}) \hat \gamma_a$ in
the cohomology basis.

On the F-theory side one has the following consistency condition on the flux. 
The first constraint comes from the quantization condition for $G_4$, 
which depends on the second Chern class of $X_4$ in the following 
way~\cite{Witten:1996md}
\begin{equation} 
G_4 + \frac{c_2(X_4)}{2} \in H^4(X_4,\mathds{Z})\ .
\label{fluxcondition1}
\end{equation}  
More restrictive is the condition that $G_4$ has to be primitive,
i.e.~orthogonal to the K\"ahler form of $X_4$. In the F-theory limit of vanishing elliptic fiber this yields the
constraints
\begin{equation} 
\int_{X_4} G_4 \wedge J_i\wedge J_j=0\ . 
\label{fluxcondition2} 
\end{equation}
for every generator $J_i$, $i=1,\ldots,h^{1,1}(X_4)$ of the K\"ahler cone.
To discuss the two conditions further it is useful to remind us of the fact 
that the  (co)homology of a Calabi-Yau splits into a horizontal 
and a vertical subspace
\begin{equation}
	H^4_H(X_4,\mathds{Z})=\bigoplus_{k=0}^4 H_H^{4-k,k}(X_4,\mathds{Z} )\,,\quad
	H^4_V(X_4,\mathds{Z})=\bigoplus_{k=0}^4 H_V^{k,k}(X_4,\mathds{Z} )\,.
\end{equation}
Since we have an even number of complex dimensions 
the group $H^{2,2}(X_4,\mathds{C})$ contains both parts and splits accordingly
into the vertical and the horizontal subspace as \cite{Greene:1993vm}
\begin{equation}
H^{2,2}(X_4)=H^{2,2}_V(X_4)\oplus  H^{2,2}_H(X_4)\ . 
\end{equation} 
Analogous to the two-dimensional case of K3 and in contrast to the Calabi-Yau threefold case, the derivatives of
$\Omega_4$ with respect to the complex structure modulo the differential ideal given by the Picard-Fuchs operators
generate only the horizontal subspace. The remaining part is the vertical subspace which is the natural 
ring of polynomials in the K\"ahler cone generators $J_i$ modulo the ideal defining the intersection ring.  
Mirror symmetry exchanges
the vertical and the horizontal subspace.  A corollary of these statements is that the allowed
fluxes in the superpotential \eqref{4ffluxsuperpotential} are in the horizontal subspace.  
On the other hand Chern classes are in
the vertical subspace, so that half integral flux quantum numbers are not
allowed if condition (\ref{fluxcondition2}) is met.
Now, the most important task on the fourfold side is to find the periods which
correspond to the integrals over an integral basis of $H_4(X_4,\mathds{Z})$.

The first step to determine the periods is to determine the 
Picard-Fuchs equations $\cL_\kappa \Pi^a(\underline{t}) = 0$ satisfied by the periods.
The Picard-Fuchs operators $\cL_\kappa$ are differential operators 
in the complex structure moduli $\underline{t}$. 
In general, the $\cL_\kappa$ can be determined by applying Griffiths-Dwork
reduction \cite{Griffiths1}.
One identifies the $\cL_\kappa$ which yield exact forms when applied to $\Omega_4$, i.e.
\beq
   \cL_\kappa \Omega_4 = d w_\kappa\ ,
\eeq 
where $w_\kappa$ are three-forms on $X_4$. To derive the Picard-Fuchs operators 
$\cL_\kappa$ one uses an explicit expression for the 
holomorphic four-form $\Omega_4$ via the Griffiths residuum expression \cite{Griffiths1}. 
For Calabi-Yau fourfold hypersurfaces and 
complete intersections $\{P_1= \ldots =P_s=0\}$ with  $dP_1\wedge\ldots \wedge dP_s\neq 0$ 
in toric varieties $\mathds{P}_\Delta$ of dimensions $s+4$ the four-form $\Omega_4$ 
can be expressed as
\begin{equation} 
\Omega_4=\int_{\epsilon_1} \ldots  \int_{\epsilon_s} \prod_{k=1}^s \frac{a_0^{(k)}}{P_k}\Delta \ . 
\label{residuumform} 
\end{equation}  
Here $\epsilon_i$ are paths in $\mathds{P}_\Delta$, which encircle $P_i=0$ and $\Delta$ 
is an measure invariant under the torus action. The parameter $a_0^{(k)}$ denotes a distinguished coefficient
in the defining constraint $P_k$ as introduced below. This method is general but tedious. 
However, the operators $\cL_\kappa$ can also be determined by the toric data. They are related to the 
scaling relations of the dual toric variety $\mathds{P}_{\tilde\Delta}$ that happens to be the ambient space of
the mirror fourfold $\tilde X_4$ of $X_4$. 
This nicely connects to the framework of toric mirror symmetry
\cite{Batyrev:1994hm,Hosono:1993qy,Berglund:1995gd} where the charge vectors 
$\ell^{(a)}$, defining the K\"ahler cone of the mirror $\tilde {X}_4$, 
determine a canonical set of differential operators, the GKZ-system, from which the Picard-Fuchs system for the complex structure of $X_4$ is 
obtained. From these operators $\cL_\kappa$ one can evaluate a finite set of solutions 
$\Pi^a(\underline{t})$.         

In a second step, one has to identify the solutions corresponding to 
the integral basis of $H_4(X_4,\mathds{Z})$. A strategy to do this 
was outlined in \cite{Grimm:2009ef} (see also
refs.~\cite{Mayr:1996sh,Klemm:1996ts}) 
and made concrete in simple examples. The key idea is to use the structure of 
the solution near conifold divisors in the moduli space, where a four-cycle $\nu$ and 
therefore the corresponding
period $\int_{\nu} \Omega_4$ vanishes. The vanishing cycle $\nu$ can often be
identified directly with generators of  $H_4(X_4,\mathds{Z})$. Associated to
each vanishing cycle, there will be a monodromy action on the period vector that is
generated by encircling the divisor in the moduli space and is patching the, in
general redundant, generators of these monodromies globally together.

Most information comes form the large complex structure, i.e.~the point of
maximal unipotent monodromy whose location is the origin in the Mori cone
coordinate system $z_a=(-1)^{\ell_0^{(a)}} \prod_{j=0}^m
a_j^{\ell_j^{(a)}}$ for a toric hypersurface $X_4$. For every entry $\ell^{(a)}_j$ of the Mori vectors $\ell^{(a)}$ 
there are parameters $a_j$ that are just the coefficients of the constraint $P_1=0$ defining $X_4$. At the point $\underline{z}=0$  several cycles $\gamma_a$ vanish and we have one analytic solution $X^0(z)=\int_{\gamma_0} \Omega_4$ and $h^{3,1}(X_4)$
logarithmic periods $X^a(z)=\int_{\gamma_a} \Omega_4 =X^0(z) \log(z_{a}) + \Sigma_a(z)$. Then
the mirror map is given by  
\begin{equation} 
t^a=\frac{X^a}{X^0}\ . 
\label{flatfourfold}
\end{equation}
Noting that $t^a\sim \log(z_a)$ at this point we can use these flat coordinates to write the leading logarithmic
structure of the period vector as
\begin{equation} 
\Pi^T=\Big(\int_{\gamma_0} \Omega_4,\ldots, \int_{\gamma_{b_H^4}} \Omega_4 \Big) 
= X^0\big(1,\, t^a,\,\tfrac{1}{2} C^{\delta}_{ab} t^a t^b,\,
\tfrac{1}{3!} C^a_{bcd} t^b t^c t^d ,\, \tfrac{1}{4!} C_{abcd} t^a t^b t^c t^d \big) \ .    
\label{4f-period} 
\end{equation}   
In particular, the grading $(\{k\})=(0,1,2,3,4)$ in powers of $t^a$ corresponds to a
grading of $\gamma_a\in H_4(X_4)$. In the complex structure given by the point
$z$ the dual cohomology group has the natural grading $H_H^4(X_4,\mathds{Z} )$.  Mirror symmetry maps this group
to the vertical cohomology $H_V(\tilde X_4,\mathds{Z})$.
Thus, the Greek indices in (\ref{4f-period})  run from $1$ to
$h_H^{2,2}(X_4)=h_V^{2,2}(\tilde X_4)$, the Latin indices from $1$ to
$h^{3,1}(X_4)=h^{1,1}(\tilde X_4)$. Note that 
we have introduced the constant coefficients $C_{ab}^\delta=\eta^{(2)\ \delta\gamma}
C^0_{ab\gamma}$, $C_{abc}^e=\eta^{(1)\ ed} C^0_{abcd}$ that are related to the classical intersection numbers $C^0_{ab\gamma}$ and $C^0_{abcd}$. These are calculated in the classical geometry of $\tilde{X}_4$ as follows. Let us denote a basis of $H_V(\tilde X_4,\mathds{Z})$
by
\beq
 A_{p_k}^{(k)}=a_{p_k}^{i_1,\ldots,i_k} \tilde J_{i_1} \wedge\ldots\wedge \tilde J_{i_k}\ ,
\eeq 
where the $\tilde J_{i_n}$ are the generators of the K\"ahler cone of the mirror $\tilde X_4$. 
Then one has 
\beq
   C^0_{abcd}=\int_{\tilde X_4}A_{a}^{(1)}\wedge A_{b}^{(1)}\wedge
A_{c}^{(1)}\wedge  A_{d}^{(1)} \ , \qquad C^0_{ab\gamma}=\int_{\tilde
X_4}A_{a}^{(1)}\wedge A_{b}^{(1)}\wedge  A_{\gamma}^{(2)}
\eeq
and  $\eta_{ab}^{(1)}=\int_{\tilde X_4}A_{a}^{(1)}\wedge A_{b}^{(3)}$ as well as $\eta_{\gamma\delta }^{(2)}= \int_{\tilde X_4}A_{\gamma}^{(2)}\wedge A_{\delta}^{(2)}$ denote subblocks of
$\eta_{ab}$ at grade $k=1$ and $k=2$ respectively whose inverses are 
indicated by upper indices. By formally replacing the $\tilde J_i$ with
$\theta_i=z_i\frac{d}{d z_i}$, we get a map 
\begin{equation}
	\xymatrix{
	\mu:H_V(\tilde X_4,\mathds{Z})\ar[r]& H^4_H( X_4,\mathds{Z})
	}
\end{equation}
given by 
\begin{equation} 
	\xymatrix{
	\mu: A_{p_k}^{(k)}\ar @{|->}[r] & \left. \left. {\cal R}_{p_k}^{(k)}\Omega_4\right|_{\underline{z}=0}:= a_{p_k}^{i_1,\ldots,i_k} \theta_{i_1}\cdots  \theta_{i_k}\Omega_4\right|_{\underline z=0},     
	}
	\label{def-cR}
\end{equation}                      
which preserves the grading. This implies that one can think of the integral basis $\hat \gamma_a$
in terms of their corresponding differential operators ${\cal R}_{p_k}^{(k)}$ acting on $\Omega_4$. 

The representation of the integral basis as differential operators 
will be particularly useful in the identification of the 
heterotic and F-theory superpotential. In particular, 
this formalism allows us to express the flux $G_4$ in an integral basis
in the form
\begin{equation} 
  G_4=\sum_{k=0}^4 \sum_{p_k} N^{p_k (k)} \left. {\cal R}_{p_k}^{(k)}\Omega_4\right|_{\underline{z}=0}\ .
\end{equation} 
In the next section we will argue that the heterotic/F-theory duality map is obtained by a matching 
of the operators $\cR_{p_k}^{(k)}$ with their analogs in the heterotic blow-up.

\subsection{Duality of the Heterotic and F-Theory Superpotentials}
\label{DualSupP}

Let us finally turn to the matching of the heterotic and F-theory 
superpotentials. Recall, that the heterotic superpotential (\ref{hetallpots}), is formally given by  
\begin{equation} 
W_{\rm het}(\underline{t}^c,\underline{t}^g,\underline{t}^o) = W_{\rm flux}(\underline{t}^c) + 
W_{\rm CS}(\underline{t}^c,\underline{t}^g) + W_{\rm M5}(\underline{t}^c,\underline{t}^o)\,,
\end{equation} 
where $t^c,t^g$ and $t^o$ denote the complex structure, bundle and five-brane
moduli respectively.
The last two terms are not inequivalent, since tuning the $\underline{t}^g$ or
$\underline{t}^o$ moduli one can condense or evaporate five-branes and
explore different branches of the heterotic moduli space. Clearly the moduli
spaces parametrized by $\underline{t}^c$ and $\underline{t}^g$ do not factorize
globally in complex structure and bundle moduli since the notion of a
holomorphic gauge bundle on $Z_3$ depends on the complex structure of $Z_3$.
Similarly, $\underline{t}^c$ and $\underline{t}^o$ do not factorize as the
notion of a holomorphic curve in $Z_3$ does depend on the complex structure of
$Z_3$. This is also reflected in the fact that flux and brane superpotential
can be unified into one superpotential \eqref{eq:relsup} for which the splitting 
into $W_{\rm {M5}}$ and $W_{\rm {flux}}$ is just a matter of basis choice of $H_3(Z_3,\cC,\mathds{Z})$.  

The key point of the construction is of course that we can map 
the heterotic moduli $(\underline{t}^c,\underline{t}^g,\underline{t}^o)$ 
to the complex structure moduli $\underline{t}$ of $X_4$ which 
are encoded in the fourfold period integrals. To make the equivalence 
\begin{equation}  \label{supermatch}
W_{\rm het}(\underline{t}^c,\underline{t}^g,\underline{t}^o)=W_{G_4}(\underline{t})\ ,
\end{equation}
precise, we need to establish a dictionary between the topological data 
on the heterotic side, which consist of the  heterotic flux quanta, 
the topological classes of gauge bundles and the class of the curves 
${\cal C}$, and the F-theory flux quanta. 

In order to study the duality map, we will restrict our considerations to 
the map between five-brane moduli and complex structure deformations 
of $Z_3$ to complex structure deformations of $\hat X_4$. This can be achieved
by restricting the heterotic gauge bundle $E$ to be of trivial $SU(1)\times SU(1)$ type.
In this case one needs to include heterotic five-branes to satisfy the 
anomaly cancellation condition \eqref{eq:anomaly}.
In accord with the discussion of section \ref{F_blowup} the
dual fourfold $\hat X_4$ can be realized as a complete intersection 
blown up along the five-brane curves. 
As above, we will restrict the discussion to a single five-brane.
We want to match this description with the heterotic theory on $\hat Z_3$.
One can now identify the blow-up constraints 
\begin{equation}
	\xymatrix{Q = l_1 g_5(\underline{u}) - l_2 \tilde{z} \ar@{|->}[r]& Q'= l_1 g_5(\underline{u}) - l_2 k} \ , \qquad \qquad \xymatrix{\tilde z \ar@{|->}[r]& k}\ ,
	\label{Q-map}
\end{equation}
where $\underline{u}$ denote coordinates on the base $B_2$, 
$\{\tilde z=0\}$ defines the base $B_2$ in $Z_3$, and $\{z=0\}\cap \{k=0\}$
defines the base $B_2$ in $X_4$.\footnote{Note that the $\P^1$-fibration 
$B_3 \rightarrow B_2$ has actually two sections. As in section \ref{F_blowup}, $k=0$ 
is one of the two sections, say, the zero section.}  
The map \eqref{Q-map} is possible since both $Z_3$ and $X_4$ share the twofold base 
$B_2$ with the curve $\cC$. The identification of $\tilde z$ with $k$
corresponds to the fact that in heterotic/F-theory duality the elliptic fibration 
of $Z_3$ is mapped to the $\P^1$-fibration of $B_3$.
Clearly, the map \eqref{Q-map} identifies the deformations of $\cC$ realized 
as coefficients in the constraint $\{Q=0\}$ of $\hat Z_3$ with the complex structure deformations of 
$\hat X_4$ realized as coefficient in $\{Q'=0\}$. We also have 
to match the remaining constraints $\{P=0\}$ and $\{P'=0\}$ of $\hat Z_3$ and 
$\hat X_4$, respectively. Clearly, there will not be a general match. 
However, as was argued in ref.~\cite{Berglund:1998ej} for Calabi-Yau fourfold hypersurfaces, 
one can split $P'=0$ as $P+ V_E$ yielding a map 
\begin{equation}
	\xymatrix{P + V_E\ar@{|->}[r]& P'}\ ,
	\label{P-map}
\end{equation}
where $V_E$ is describing the spectral cover of the dual heterotic bundles $E=E_1 \oplus E_2$.
Again, this requires an identification of $\tilde z$ and $k$. For $SU(1)$ bundles this map 
was given in \eqref{Q-map}, but can be generalized for non-trivial bundles.
Note that the maps \eqref{Q-map} and \eqref{P-map} can also be formulated in terms of the GKZ systems 
of the complete intersections $\hat Z_3$ and $\hat X_4$. It implies that the $\ell^{(a)}_i$ of $\hat X_4$
contain the GKZ system of $Z_3$ and the five-brane $\ell$-vectors, similar to the situation encountered in refs.~\cite{Alim:2009rf,Alim:2009bx,Grimm:2009ef,Aganagic:2009jq,Li:2009dz}.

To match the superpotentials as in \eqref{supermatch} one finally has to identify 
the integral basis of $H^{3}(\hat Z_3,D,\Z)$ with elements of $H^{4}(\hat X_4,\Z)$
and show that the relative periods of $\hat \Omega_3$ can 
be identified with a subset of the periods of $\Omega_4$. In order 
to do that, one compares the residue integrals \eqref{ResZhat}  and \eqref{residuumform}
for $\hat \Omega_3$ and $\Omega_4$ represented as complete intersections. 
Using the maps \eqref{Q-map} and \eqref{P-map} one then shows that 
each Picard-Fuchs operator annihilating $\hat \Omega_3$ is also annihilating 
$\Omega_4$. Hence, also a subset of the solutions to the Picard-Fuchs equations can 
be matched accordingly. As a minimal check, one finds that the periods of $\Omega_3$ on 
$Z_3$ before the blow-up arise as a subset of the periods of $\Omega_4$ is specific directions \cite{Mayr:1996sh,Grimm:2009ef}. The map between the cohomologies 
$H^{3}(\hat Z_3,D,\Z) \hookrightarrow H^{4}(\hat X_4,\Z)$ is 
also best formulated in terms of operators $\cR_p^{(i)}$ applied to 
the forms $\hat \Omega$ and $\Omega_4$, 
\begin{equation} 
	\xymatrix{
\left. {\cal R}_{p}^{(i)}\tilde \Omega_3(\underline{z}^c,\underline{z}^o)\right|_{\underline{z}^{c}=\underline{z}^{o}=0}\quad
\ar@{|->}[r]& \quad \left. {\cal R}_{p}^{(i)}\Omega_4(\underline{z})\right|_{\underline z=0}
	}\ .
\end{equation}
Note that the preimage of this map will in general contain derivatives with
respect to the variables $z^o$ and hence is an element in relative cohomology.
It was shown in refs.~\cite{Jockers} that one can find differential operators
$\cR^{(i)}_p$ which span the full space $H^{3}(\hat Z_3,D,\Z)$. One now finds
that by identifying the heterotic and F-theory moduli at the large complex
structure point $\underline{z}=0$, one obtains an embedding map of the integral
basis.

One immediate application of this formalism is that if we know the 
classical quadratic terms in $W_{\rm het}$ we can fix  the dual $G_4$-flux and 
use the periods of the fourfold to determine the instanton parts. In particular, 
for the five-brane superpotential $W_{\hat M5}(t^o,t^c)$ one finds 
that the dual flux $G^{\rm M5}_4$ can be expressed as
\begin{equation} 
G_4^{\rm M5}=\sum_{p} N^{p (2)} \left. {\cal R}_{p}^{(2)}\Omega_4\right|_{\underline z=0}
\end{equation}  
Note that for $G_4$ fluxes generated by operators $\cR^{(2)}$ the superpotential yields an
integral structure of the fourfold symplectic invariants at large
volume of the mirror Mir$(\hat X_4)$ of $\hat X_4$ as~\cite{Greene:1993vm,Mayr:1996sh,Klemm:2007in}
\beq
\label{Li2}
   W_{G_4}^{\rm inst}=\sum_{\beta \in H_2(Mir(\hat X_4),\Z)} n^0_\beta(\gamma_{G_4}) {\rm Li}_2(\underline{q}^\beta)\ ,\quad n^0_\beta\in\Z\ ,
\eeq 
where $\gamma_{G_4}$ is co-dimension two cycle specified
by the flux~\cite{Grimm:2009ef}, and ${\underline{q}}^\beta= \text{exp}({\int_{\beta} \tilde J})$ is 
the exponential of the mirror K\"ahler form $\tilde J$ integrated over classes $\beta$. 
This integrality structure is inherited to the heterotic superpotentials in geometric
phases of their parameter spaces.  For superpotential from  five-branes wrapped
on a curve ${\cal C}$  this matches naturally the disk multi-covering formula
of~\cite{Ooguri:1999bv}, since this part is mapped by mirror symmetry to disk
instantons ending on special Lagrangians $L$ mirror dual to ${\cal C}$. It would 
be interesting to explore a generalization of this integral structure to 
the gauge sector of the heterotic theory.

Finally, there is geometric way to identify the flux which corresponds to a
chain integral $\int_{\Gamma} \Omega_3$. The three-chain ${\Gamma}$ can be
mapped to a three-chain  ${\Gamma}$ in $B_3$ whose boundary two-cycles lie in
the worldvolume of a seven-brane over which  the cycles of the F-theory
elliptic fiber degenerates.  By fibering the one-cycle of the elliptic fiber which
vanishes at the seven-brane locus over $\Gamma$, one gets a transcendental
cycle in $H_4(X_4,\mathds{Z})$. Its dual form lies then in the horizontal part
$H^4_H(X_4,\mathds{Z})$ and therefore yields the flux (see
ref.~\cite{Denef:2008wq} for a review on such constructions).  For a recent
very explicit construction of these cycles in F-theory compactifications on
elliptic K3 surfaces and Calabi-Yau threefolds see refs.~\cite{Braun:2008ua}.

\section{Examples of Heterotic/F-Theory Dual Pairs}
\label{OurExamples}

In this section we study concrete examples to demonstrate the
concepts discussed in the earlier sections. We will examine 
two geometries in detail. The first F-theory Calabi-Yau fourfold geometry, discussed in section \ref{Example1} and 
\ref{non-CYblowup}, will have few K\"ahler moduli and many complex structure
moduli. In this case we can use toric geometry to 
compute explicitly the intersection numbers, evaluate both sides of the expression 
\eqref{eq:modulimap} yielding the number of deformation moduli of the five-brane curve, and check 
the anomaly formula \eqref{eq:anomaly}. We also show that the Calabi-Yau fourfold 
can be explicitly constructed from the heterotic non-Calabi-Yau threefold obtained 
by blowing up the five-brane curve in section \ref{non-CYblowup}.
The second Calabi-Yau fourfold example, introduced in section \ref{Example2}, 
will admit few complex structure moduli
and many K\"ahler moduli. This allows us to identify the bundle
moduli and five-brane moduli under duality by studying the Weierstrass constraint.
The F-theory flux superpotential for this configuration was already evaluated in
ref.~\cite{Grimm:2009ef}, and we will discuss its heterotic dual in section
\ref{Example2}.

\subsection{Example 1: Five-Branes in the Elliptic Fibration over $\P^2$}
\label{Example1}

We begin the discussion of our first example of heterotic/F-theory dual theories by defining the setup on the heterotic side.
Following section \ref{hettransition} the heterotic theory is specified by an elliptic Calabi-Yau threefold $Z_3$ with a stable holomorphic vector bundle $E=E_1\oplus E_2$ obeying the heterotic anomaly constraint \eqref{eq:anomaly}.

We choose the threefold $Z_3$ as the elliptic fibration 
over the base $B_2=\P^2$ with generic torus fiber $\P_{1,2,3}[6]$. It is given as a hypersurface $P=0$ in the toric ambient space
\begin{equation}
\label{Z3poly}
	\centering
	\Delta(Z_3)=
	\left(
	\begin{tabular}{cccc|c}
		-1 & 0 & 0 & 0 & $3B+9H$\\
		0 & -1 & 0 & 0 & $2B+6H$\\
		3 & 2& 0 & 0 & $B$\\
		3 & 2 & 1 & 1  &$H$\\
		3 & 2 & -1 & 0&$H$\\
		3 & 2 & 0 & -1&$H$
	\end{tabular}
	\right)
\end{equation}
with the class of the hypersurface $Z_3$ given by 
\begin{equation}
	[Z_3]=\sum D_i=6B+18H\,.
\end{equation} 
Here we denoted the two independent toric divisors $D_i$ by $H$ and $B$, the pullback of the hyperplane class of the $\mathds{P}^2$ base respectively the class of the base itself. 
From the toric data the basic topological numbers of $Z_3$ are obtained as
\beq
	\chi(Z_3)=540\ ,\quad h^{1,1}(Z_3)=2\ ,\quad h^{2,1}(Z_3)=272\ .
	\label{eqn:top-data-tfp2}
\eeq
The second Chern-class of $Z_3$ is in general given in terms of the Chern classes $c_1(B_2)$, $c_2(B_2)$ and the section $\sigma:B_2\rightarrow Z_3$ of the elliptic fibration by $c_2(Z_3)=12 c_1(B_2)\sigma+11 c_1(B_2)^2+c_2(B_2)$. Here we have $\sigma=B$ and thus obtain 
\begin{equation}
 	c_2(Z_3)=36H\cdot B+102 H^2\,.
\end{equation}

To satisfy the heterotic anomaly formula \eqref{eq:anomaly}, we 
have to construct the heterotic vector bundle $E_1 \oplus E_2$ and 
compute the characteristic classes $\lambda(E_i)$.  Since $Z_3$ is elliptically fibered the classes $\lambda(E_i)$ 
can be constructed using the basic methods of \cite{Friedman:1997yq} that were briefly mentioned in section \ref{F_basics}. 
According to \eqref{eq:c2E}, we first need to specify the classes $\eta_1,\eta_2\in H^{2}(B_2,\Z)$ essential in the
spectral cover construction. We furthermore
restrict $E_1\oplus E_2$ to be an $E_8\times E_8$ bundle over $Z_3$ and choose both classes as $\eta_1 = \eta_2 = 6c_1(B_2)$. 
Then, we use the formula for the second Chern class of $E_8$-bundles 
\beq
	\lambda(E_i)=\frac{c_2(E_i)}{60}=\eta_i\sigma-15\sigma^2+135\eta_i
	c_1(B_2)-310c_1(B_2)^2\,
	\label{eqn:lambda-e8}
\eeq
to obtain $\lambda(E_1)=\lambda(E_2)=18 H\cdot B-360 H^2$.
The anomaly condition \eqref{eq:anomaly} then leads to conditions on the coefficients of the independent classes in $H^4(Z_3)$.
For the class $H\cdot B$ contributed by the base via $\sigma\cdot H^2(\mathds{P}^2,\Z)$ this is trivially satisfied by the choice of $\lambda(E_i)$. This implies that no horizontal five-branes are present. 
For the class of the fiber $F$ the anomaly forces the inclusion of vertical five-branes in the class $\mathcal{C}=c_2(B_2)+91c_1(B_2)^2=822 H^2 \equiv n_f F$.
Since $F$ is dual to the base $B_2$ the number $n_f$ of vertical branes is determined by integrating $\mathcal{C}$ over $\mathds{P}^2$, 
\beq
	n_f=\int_{\mathds{P}^2}\cC=822.
	\label{eqn:nf}
\eeq
To conclude the heterotic side we compute the index $I(E_i)$ since it appears
in the identification of moduli \eqref{eq:modulimap} and thus is crucial for
the analysis of heterotic/F-theory duality. For $Z_3$ we use the formula
\eqref{eqn:index} to obtain that $I(E_1)=I(E_2)=8+4\cdot 360+18\cdot3=1502$.   

Next we include horizontal five-branes to the setup by shifting the classes $\eta_i$ appropriately. 
We achieve this by putting $\eta_2=6c_1(B)-H$. The 
class of the five-brane $\mathcal C$ can then be determined analogous to the above discussion by evaluating \eqref{eqn:lambda-e8} and imposing the anomaly \eqref{eq:anomaly}. It takes the form
\begin{equation}
 	\cC=91 c_1(B_2)^2 + c_2(B_2) - 45 c_1(B_2)\cdot H + 15 H^2 + H\cdot B=702 H^2+H\cdot B\,,
\end{equation}
which means that we have to include five-branes in the base on a curve $\cC$ in the class $H$ of the hyperplane of $\mathds{P}^2$.
Additionally the number of five-branes on the fiber $F$ is altered to $n_f=702$.
Accordingly, the shifting of $\eta_2$ changes the second index to $I_2=1019$, whereas $I_1=1502$ remains unchanged.

Let us now turn to the dual F-theory description. We first construct the
fourfold $X_4$ dual to the heterotic setup with no five-branes. In this case
the base $B_3$ of the elliptically fibered fourfold is $B_3=\P^1\times \P^2$.
This can be seen from the relation \eqref{eq:etaE8} of the classes $\eta_i$ and
the fibration structure of $B_3$ for $E_8$-bundles. Since both classes equal
$6c_1$ we have $t=0$ and thus the bundle $L=\mathcal{O}_{\mathds{P}^2} $ is
trivial as well as the projective bundle
$B_3=\mathds{P}(\mathcal{O}_{\mathds{P}^2}\oplus\mathcal{O}_{\mathds{P}^2})$.
Then the fourfold $X_4$ is constructed as the elliptic fibration over $B_3$
with generic fiber given by $\mathds{P}_{1,2,3}[6]$.  Again $X_4$ is described
as a hypersurface in a five-dimensional toric ambient space $V_5$ as described
by the toric data in \eqref{eqn:fourfold1} if one drops the point
$(3,2,-1,0,1)$ and sets the divisor $D$ to zero. The class of $X_4$ is then
given by 
\begin{equation}
	\left[X_4\right]= \sum_i D_i = 6 B + 18 H + 12 K\ , 
\label{eq:X4}
\end{equation}
where the independent divisors are the base $B_3$ denoted by $B$, the pullback
of the hyperplane $H$ in $\mathds{P}^2$ and of the hyperplane $K$ in
$\mathds{P}^1$.  Then, the basic topological data reads  
\beq
	\chi(X_4)=19728\quad,\quad h^{1,1}(X_4)=3\quad ,\quad h^{3,1}(X_4)=3277\quad ,\quad h^{2,1}(X_4)=0.
	\label{eqn:top-data-ffp2}
\eeq

Now we have everything at hand to discuss heterotic/F-theory duality along the
lines of section \ref{F_blowup}, in particular the map of moduli
\eqref{eq:modulimap}. As discussed there, the complex structure moduli of the
F-theory fourfold are expected to contain the complex structure moduli of $Z_3$
on the heterotic side as well as the bundle and brane moduli of possible
horizontal five-branes. Indeed we obtain a complete matching by adding up all
contributions in \eqref{eq:modulimap},
\begin{equation}
	h^{3,1}(X_4)=3277=272+1502+1502+1\,,
\end{equation}
where it is crucial that no horizontal five-branes with possible brane moduli
are are present. 

To obtain the F-theory dual of the heterotic theory with horizontal
five-branes, we have to apply the recipe discussed in section \ref{F_blowup}.
We have to perform the described geometric transition of first tuning the
complex structure moduli of the fourfold $X_4$ such that it becomes singular
over the curve $\cC$ which we then blow up into a divisor $D$. This way we
obtain a new smooth Calabi-Yau fourfold denoted by $\hat X_4$.  The toric data
of this fourfold are given by
\begin{equation}
	\Delta(\hat X_4)=\left(
	\begin{array}{ccccc|c|c}
		-1 &  0 &  0  & 0  & 0    & 3D+3B+9H+6K & D_1\\
		0  & -1 &  0  & 0  & 0    & 2D+2B+6H+4K & D_2\\
		3  &  2 &  0  & 0  & 0    & B & D_3\\
		3  &  2 &  1  & 1  & 0    & H & D_4\\
		3  &  2 &  -1 & 0  & 0    & H-D & D_5\\
		3  &  2 &  0  & -1 & 0    & H & D_5\\
		3  &  2 &  0  & 0  & 1    & K & D_7\\
		3  &  2 &  0  & 0  & -1    & K+D & D_8\\ 
		3  &  2 &  -1  & 0  & 1   & D & D_9\\
	\end{array}
	\right).
	\label{eqn:fourfold1}
\end{equation}
where we included the last point $(3,2,-1,0,1)$ and a corresponding divisor
$D_9=D$ to perform the blow-up along the curve $\cC$ as follows. 

Since the curve $\cC$ on the heterotic theory is in the class $H$ we have to
blow-up over the hyperplane class of $\mathds{P}^2$ in $B_3$. First we project
the polyhedron $\Delta(\hat{X}_4)$ to the base $B_3$ which is done just by
omitting the first and second column in \eqref{eqn:fourfold1}. Then the last
point maps to the point $(-1,0,1)$ that subdivides the two-dimensional cone
spanned by $(-1,0,0)$ and $(0,0,1)$ in the polyhedron of $B_3$. Thus, upon
adding this point the curve $\cC=H$ in $B_2$ corresponding to this cone is
removed from $B_3$ and replaced by the divisor $D$ corresponding to the new
point. Thus we see that the toric data \eqref{eqn:fourfold1} contain this
blown-up base $B_3$ in the last three columns.

The fourfold is then realized as a generic constraint $P = 0$ in the class
\beq
  [\hat{X}_4] = 6 B + 18 H + 12 K + 6 D\ .
\eeq  
Note that this fourfold has now three different triangulations which correspond to 
the various five-brane phases on the dual heterotic side.
The topological data for the new fourfold $\hat X_4$ are given by
\beq
	\chi(\hat X_4)=16848\quad,\quad h^{1,1}(\hat X_4)=4\quad,\quad h^{3,1}(\hat X_4)=2796\quad ,\quad h^{2,1}(\hat X_4)=0\ ,
	\label{eqn:xhat-top-data}
\eeq
where the number of complex structure moduli has reduced in the transition as
expected.

If we now analyze the map of moduli \eqref{eq:modulimap} in heterotic/F-theory
duality we observe that we have to put $h^0(\mathcal{C},N_{Z_3}\mathcal C)=2$
in order to obtain a matching.  This implies, from the point of view of
heterotic/F-theory duality, that the horizontal five-brane wrapped on $\cC$ has
to have two deformation moduli. Indeed, this precisely matches the fact that
the hyperplane class of $\P^2$ has two deformations since a general hyperplane
is given by the linear constraint $a_1x_1+a_2x_2+a_3x_3=0$ in the three
homogeneous coordinates $x_i$ of $\mathds{P}^2$. Upon the overall scaling it
thus has two moduli parameterized by the $\mathds{P}^2$ with homogeneous
coordinates $a_i$. This way we have found an explicit construction of an
F-theory fourfold with complex structure moduli encoding the dynamics of
heterotic five-branes.

In section \ref{non-CYblowup} we provide further evidence for this
identification by showing that one can also construct $\hat X_4$ as a complete
intersection starting with a heterotic non-Calabi-Yau threefold.
Unfortunately, it will be very hard to compute the complete superpotential for
the fourfold $\hat X_4$ since it admits such a large number of complex
structure deformations.  It would be interesting, however, to extract the
superpotential for a subsector of the moduli including the two brane
deformations.\footnote{If one considers exactly the mirror of $\hat X_4$, as we
will in fact do in section \ref{Example2}, it might be possible to embed this
reduced deformation problem into the complicated deformation problem of $\hat
X_4$ constructed in this section.} Later on we will take a different route and
consider examples with only a few complex structure moduli which are
constructed by using mirror symmetry.

\subsection{Calabi-Yau Fourfolds from Heterotic Non-Calabi-Yau Threefolds}
\label{non-CYblowup}

In this section we discuss the example of section \ref{Example1} 
employing the blow-up proposal of ref.~\cite{Grimm:2008dq} as 
discussed in section~\ref{heterotic_blowup}.
More precisely, we will explicitly construct a non-Calabi-Yau 
threefold $\hat Z_3$ which is obtained by blowing up the horizontal five-brane 
curve into a divisor. This translates the deformations of $\cC$ into new
complex structure deformations of $\hat Z_3$. The F-theory Calabi-Yau 
fourfold $\hat X_4$ is then naturally obtained from the base of $\hat Z_3$ by an 
additional $\P^1$ fibration. $\hat X_4$ is identical to the fourfold considered 
in section \ref{Example1}, despite the fact that it is now realized as a 
complete intersection.

As in section \ref{Example1} the starting point is the elliptic 
fibration $Z_3$ over $B_2=\P^2$  with a five-brane wrapping the 
hyperplane class of the base.
Let us describe the explicit construction of $\hat Z_3$.
The blow-up geometry $\hat Z_3$ is given by $\P(N_{Z_3}\mathcal C)$.
$Z_3$ is a hypersurface $\{P=0\}$ in a toric variety $V_4$ and 
the curve $\mathcal C$ is given as a complete intersection of two hypersurfaces in $Z_3$, i.e.~${\mathcal C}=\{h_1=0\}\cap\{h_2=0\}\subset Z_3$.
The charge vectors of $V_4$ are given by $\{\ell^{(i)}\}$ with $i=1,\dots,k$.
We are aiming to construct a five-dimensional toric variety which is given by $\hat V_5=\P(N_{V_4}\mathcal C)$ and 
use the blow-up equation described in section~\ref{heterotic_blowup}.
Let us denote the divisor classes defined by $h_i$ by $H_i$ and the charges of $h_i$ by $\mu_i=(\mu_{i}^{(1)},\dots\mu_{i}^{(k)})$.
Then, the coordinates $l_i$ of $N_{V_4}H_i$ transform with charge $\mu_i^{(m)}$
under the $k$ scaling relations.
The normal bundle $N_{V_4}\mathcal C$ is given by $N_{V_4} H_1\oplus N_{V_4} H_2$.
Since we have to projectivize $N_{V_4} \cC$, we have to include another $\C^*$-action with charge vector $\ell^{(k+1)}_{\hat V_5}$ acting non-trivially only 
on the new coordinates $l_i$.
The new charge vectors of $\hat V_5$ are thus given by the following table
\begin{center}
	\begin{tabular}{|c||c|c|c|}
		\hline\T  & coordinates of $V_4$ & $l_1$ & $l_2$ \B\\
		 \hline  $\ell^{(1)}_{\hat V_5}$ & $\ell^{(1)}$ & $\mu_{1}^{(1)}$ & $\mu_{2}^{(1)}$ \B\\
		 \hline\T \vdots & \vdots & \vdots & \vdots \B\\
		 \hline\T $\ell^{(k)}_{\hat V_5}$ & $\ell^{(k)}$ & $\mu_{1}^{(k)}$ & $\mu_{2}^{(k)}$\B \\
		 \hline\T $\ell^{(k+1)}_{\hat V_5}$ & 0 & 1 & 1 \B\\ \hline
	\end{tabular}
\end{center}

As in \eqref{eq:blowup}, the blown-up geometry $\hat Z_3$ is now given as a complete intersection
\beq
	P=0\quad,\quad l_1 h_2 -l_2 h_1 =0\ .
	\label{eqn:blowup-complete-intersection-eq}
\eeq
To apply this to the elliptic fibration over $\P^2$ with the polyhedron \eqref{Z3poly}, one 
picks the curve $\mathcal C$ given by $\{\tilde{z}=0\}$ and $\{x_1=0\}$.
$\cC$ is a genus zero curve and we will find that the exceptional divisor $D$ will be the first 
del Pezzo surface $dP_1$ in accord with the discussion of section \ref{heterotic_blowup}.
We construct the five-dimensional ambient manifold as explained above,
\begin{equation}
	\centering
	\Delta(\hat Z_3)=\left(
	\begin{tabular}{ccccc|c}
		-1 & 0 & 0 & 0& 0 & $3B+3D+9H$\\
		0 & -1 & 0 & 0 & 0 &$2B+2D+6H$\\
		3 & 2& 0 & 0 & 0 & $B$\\
		3 & 2 & 1 & 1 & 1 & $H$\\
		3 & 2 & -1 & 0 & 0 & $H$\\
		3 & 2 & 0 & -1 & 0 & $H$\\
		3 & 2 & 0 & 0 & -1 & $D$\\
		0 & 0 & 0 & 0 & -1 & $H-D$
	\end{tabular}
	\right).
	\label{eqn:blowup-poly}
\end{equation}
Note that one has to include the inner point $(3,2,0,0,0)$ 
which corresponds to the base of the elliptic fibration $\hat Z_3$.
Furthermore, one shows that the point $(0,0,0,0,1)$, required for the above 
scalings, can be omitted since the associated divisor does not intersect the complete intersection 
$\hat Z_3$. Explicitly the complete intersection $\hat Z_3$ is given by a generic constraint in 
the class 
\beq
	\hat Z_3: \quad (6B+6D+18H) \cap H\ ,
	\label{eqn:ci-blow-up}
\eeq
where $H,B,D$  are the divisor classes of the ambient space \eqref{eqn:blowup-poly}.
The first divisor in \eqref{eqn:ci-blow-up} is the sum of the first seven divisors in \eqref{eqn:blowup-poly} 
and corresponds to the original Calabi-Yau constraint $P=0$ in
\eqref{eqn:blowup-complete-intersection-eq}.  The second divisor in \eqref{eqn:ci-blow-up} is the sum of
the last two divisors and is the class of the second equation of 
\eqref{eqn:blowup-complete-intersection-eq}.  This complete intersection
threefold has $\chi(\hat Z_3)=-538=\chi(Z_3)-\chi(\P^1)+\chi(dP_1)$, and one checks that the exceptional 
divisor $D$ has the characteristic data of a del Pezzo $1$ surface. 
This means that we have replaced the hyperplane isomorphic to $\P^1$ in the base with the exceptional divisor 
which is $dP_1$. It can be readily checked that the first Chern class of $\hat Z_3$ is 
non-vanishing and equals $-D$.

Having described the blow-up geometry, we now turn to the construction
of the fourfold $\hat X_4$ for F-theory.
This fourfold will also be constructed as complete intersection, but it will
be the same manifold as the fourfold described in section \ref{Example1}, equation \eqref{eqn:fourfold1}.
We fiber an additional $\P^1$ over
$\P(\Delta(\hat Z_3))$ which is only non-trivially fibered along the exceptional
divisor.
This is analogous to the construction of the dual fourfold in the heterotic/F-theory duality 
where one also fibers $\P^1$ over the base twofold of the Calabi-Yau threefold to obtain the F-theory fourfold.
Here we proceed in a similar fashion but construct a $\P^1$-fibration over the base of the 
non-Calabi-Yau manifold $\hat Z_3$. This base is a complete intersection and thus leads to a realization of  
$\hat X_4$ as a complete intersection. Concretely, we have the following polyhedron
\begin{equation}
	\Delta(\hat X_4)=\left(
	\begin{array}{cccccc|c|c}
		-1 &  0 &  0  & 0  & 0  & 0  & 3D+3B+9H+6K & D_1 \\
		0  & -1 &  0  & 0  & 0  & 0  & 2D+2B+6H+4K & D_2  \\
		3  &  2 &  0  & 0  & 0  & 0  & B & D_3 \\
		3  &  2 &  1  & 1  & 1  & 0  & H& D_4 \\
		3  &  2 &  -1 & 0  & 0  & 0  & H& D_5 \\
		3  &  2 &  0  & -1 & 0  & 0  & H& D_6 \\
		3  &  2 &  0  & 0  & 0  & 1  & K& D_7 \\
		3  &  2 &  0  & 0  & 0  & -1 & K+D& D_8 \\ \hline
		0  &  0 &  0  & 0  & -1 & 1  & D& D_9 \\
		0  &  0 &  0  & 0  & -1  & 0 & H-D& D_{10} 
	\end{array}
	\right).
	\label{eqn:ci-fourfoldpoly}
\end{equation}
The fourfold $\hat X_4$ is given as the following complete intersection
\beq
	\hat X_4:\quad (6B+6D+18H+12K)\cap H\ .
	\label{eqn:ci-fourfold}
\eeq
Note that this fourfold is indeed Calabi-Yau as can be checked explicitly by analyzing the toric data \eqref{eqn:ci-fourfoldpoly}. 
For complete intersections the Calabi-Yau constraint is realized via the two partitions, so-called nef partitions, 
in \eqref{eqn:ci-fourfoldpoly} as in refs.~\cite{Batyrev:1994pg}. The first nef partition yields the sum of the first eight 
divisors $\sum_{i=1}^{8}D_i$ in \eqref{eqn:ci-fourfoldpoly} and gives the 
first constraint in \eqref{eqn:ci-fourfold}. The second nef partition yields the sum of the last 
two divisors $D_9+D_{10}$ in \eqref{eqn:ci-fourfoldpoly}
and yield the second constraint in  \eqref{eqn:ci-fourfold}.
The divisors $D_7$ and $D_8$ correspond to the $\P^1$ fiber in the 
base of $\hat X_4$ obtained by dropping the first two columns in 
\eqref{eqn:ci-fourfoldpoly}. This fibration is only non-trivial over the 
exceptional divisors $D_9=D$ in the second nef partition of \eqref{eqn:ci-fourfoldpoly}. 
Note that if one simply drops $K$ from the expression \eqref{eqn:ci-fourfold}
one formally recovers the constraint \eqref{eqn:ci-blow-up} of $\hat Z_3$.
To check that the complete intersection $\hat X_4$ is precisely the 
fourfold constructed in section \ref{Example1}, one has to compute the 
intersection ring and Chern classes. In particular, it is not hard to show that 
also \eqref{eqn:ci-fourfoldpoly} has three triangulations matching the 
result of section \ref{Example1}.

In summary, we have found that there is a natural construction of $\hat X_4$ as
complete intersection with the base obtained from the heterotic non-Calabi-Yau threefold 
$\hat Z_3$. Let us stress that this construction will straightforwardly generalize to dual 
heterotic/F-theory setups with other toric base spaces $B_2$ and different types 
of bundles. For example, to study the bundle configurations on $Z_3$ of section \ref{Example1} 
with $\eta_{1,2} = 6 c_1(B_2) \pm k H,\ k=0,1,2$ one has 
to replace 
\beq
  D_4 \rightarrow (3,2,1,1,k)\ ,\qquad D_4  \rightarrow  (3,2,1,1,1,k)\ ,
\eeq
in the polyhedra \eqref{eqn:fourfold1} and \eqref{eqn:ci-fourfoldpoly}, respectively.
Moreover, also bundles
which are not of the type $E_8 \times E_8$ can be included by generalizing the form 
of the $\P^1$ fibration just as in the standard construction of dual F-theory fourfolds.

\subsection{Example 2: Five-Brane Superpotential in Heterotic/F-Theory Duality}
\label{Example2}

Let us now discuss a second example for which the F-theory flux superpotential 
can be computed explicitly since the F-theory fourfold 
admits only few complex structure moduli.
Clearly, using mirror symmetry such fourfolds can be obtained 
as mirror manifolds of examples with few K\"ahler moduli.
To start with, let us consider heterotic string theory 
on the \textit{mirror} of the Calabi-Yau threefold which is 
an elliptic fibration over $\P^2$. This mirror is 
the heterotic manifold $Z_3$. One shows by using the methods of 
ref.~\cite{Avram:1996pj}, that this $Z_3$ is also elliptically fibered, such 
that, at least in principle, one can construct the bundles explicitly. 
The polyhedron of $Z_3$ is the dual polyhedron to \eqref{Z3poly} and the Weierstrass form of $Z_3$ is as follows
\begin{equation}
	\mu_3=x^3+y^2+x y \tilde{z} a_0 u_1 u_2 u_3+\tilde{z}^6( a_1 u_1^{18}+ a_2 u_2^{18}+ a_3 u_3^{18}+ a_4 u_1^6 u_2^6 u_3^6).
	\label{eqn:exp2-het-threefold}
\end{equation}
The coordinates $\{u_i\}$ are the homogeneous coordinates of the twofold base $B_2$. 
Note that one finds that the elliptic fibration of this $Z_3$ is highly degenerate over $B_2$. 
The threefold is nevertheless non-singular 
since the singularities are blown up by many divisors in the toric ambient space of $Z_3$. 
In writing \eqref{eqn:exp2-het-threefold} many of the coordinates parameterizing 
these additional divisors have been set to one.\footnote{Note that the blow-down of these 
divisors induces a large non-perturbative gauge group in the heterotic compactification.}
Turning to the perturbative gauge bundle $E_1 \oplus E_2$ we will restrict in 
the following to the simplest bundle $SU(1)\times SU(1)$ which thus preserves the 
full perturbative $E_8 \times E_8$ gauge symmetry in four dimensions. 
To nevertheless satisfy the anomaly condition \eqref{eq:anomaly} one also has to include 
five-branes. In particular, we consider a five-brane in  
$Z_3$ given by the equations
\begin{equation}
	h_1 = b_1u_1^{18}+b_2u_1^6u_2^6u_3^6=0\ , \qquad h_2 = \tilde{z} =0\ .
	\label{eqn:exp2-brane-heterotic}
\end{equation}
The curve $\cC$ wrapped by the five-brane is thus in the base $B_2$ of $Z_3$.
Unfortunately, it is hard to check \eqref{eq:anomaly} explicitly as in the 
example of section \ref{Example1} since there are too many K\"ahler classes 
in $Z_3$. However, one can proceed to construct the associated 
Calabi-Yau fourfold $\hat X_4$ which encodes a consistent completion of the setup.

The associated fourfold $\hat X_4$ cannot be constructed as it was done in section~\ref{Example1}. 
However, one can employ mirror symmetry to first obtain the mirror fourfold Mir$(\hat X_4)$ of $\hat X_4$
as Calabi-Yau fibration 
\begin{equation}
	\xymatrix{
	\text{Mir}(Z_3) \ar[r] & 	\text{Mir}(\hat X_4) \ar[d]\\
	& \P^1
	}
\end{equation}
where Mir($Z_3$) is the mirror of the heterotic threefold $Z_3$ \cite{Berglund:1998ej}.
This naturally leads us to identify $\hat X_4$ as the mirror to the fourfold \eqref{eqn:fourfold1}
from section~\ref{Example1}. This fourfold is also the main example discussed in 
detail in ref.~\cite{Grimm:2009ef}.
In the following we will check that this is indeed the correct identification by using the 
formalism of refs.~\cite{Berglund:1998ej,Diaconescu:1999it}.
The Weierstrass form of $\hat X_4$ can be computed using the dual polyhedron of \eqref{eqn:fourfold1} yielding
\begin{equation} \label{eqn:weierstrass-x4}
	\mu_4=y^2+ x^3+m_1 (u_i,w_j,k_m) x y z+	m_6 (u_i,w_j,k_m) z^6=0\,,
\end{equation}
where
\bea  \label{eqn:def-m1-m6}
\nonumber 	m_1(w_j,u_i)&=&a_{0} u_1 u_2 u_3 w_1 w_2 w_3 w_4 w_5 w_6 k_1 k_2\,,\\ 
\nonumber 	m_6 (w_j,u_i)&=&\phantom{+} a_1 (k_1 k_2)^6 u_1^{18} w_1^{18} w_2^{18} w_5^6 w_6^6+a_2 (k_1 k_2)^6 u_2^{18} w_3^{18} w_5^{12}\\
	&\phantom=&+a_3 (k_1 k_2)^6 u_3^{18} w_4^{18} w_6^{12}+ a_4 (k_1 k_2)^6 (u_1 u_2 u_3 w_1 w_2 w_3 w_4 w_5 w_6)^6\\
\nonumber	&\phantom=&+ b_1 k_2^{12} u_1^{18} w_1^{24} w_2^{12} w_3^6 w_4^6+b_2 k_2^{12} (u_1u_2u_3)^6 (w_1 w_3 w_4)^{12}\\
\nonumber	&\phantom =&+c_1 k_1^{12} (u_1 u_2 u_3)^6 (w_2 w_5 w_6)^{12}.
\eea
The coordinates $u_i$ are the coordinates of the base twofold $B_2$ as before
and $w_i,k_1,k_2$ are the additional coordinates of the base threefold $B_3$. Again,
note that we have set many coordinates to one to display $\mu_4$. The chosen
coordinates correspond to divisors which include the vertices of $\Delta(X_4)$
and hence determine the polyhedron fully.  In particular, one finds that
$k_1,k_2$ are the coordinates of the fiber $\P^1$ over $B_2$.  The coefficients
$a_i,b_1,b_2,c_1$ denote coefficients encoding the complex structure
deformations of $\hat X_4$.  However, since $h^{3,1}(\hat X_4)=4$, there are
only four complex structure parameters rendering six of the $a_i$ redundant.  As
the first check that $\hat X_4$ is indeed the correct geometry, we use the
stable degeneration limit \cite{Friedman:1997yq} and write $\mu_4$ in a local
patch with appropriate coordinate redefinition as follows \cite{Berglund:1998ej}
\beq
	\mu_4=p_0+p_++p_-\, ,
\eeq
where
\bea \label{def-p0p+p-}
	\nn p_0&=& x^3+y^2+x y \tilde{z} a_0 u_1 u_2 u_3+\tilde{z}^6\left( a_1 u_1^{18}+ a_2 u_2^{18}+ a_{3} u_3^{18} + a_4 u_1^6 u_2^6 u_3^6\right)\ ,\\
	p_+&=& v\tilde{z}^6 \left(b_1 u_1^{18}+b_2 u_1^6 u_2^6 u_3^6\right)\ ,\\
	\nn p_-&=& v^{-1}\tilde{z}^6 c_1 u_1^6 u_2^6 u_3^6\ .
\eea
The coordinate $v$ is the affine coordinate of the fiber $\P^1$.  In the stable
degeneration limit $\{p_0=0\}$ describes the Calabi-Yau threefold of the
heterotic string.  In this case $p_0$ coincides with $\mu_3$ which means that
the heterotic Calabi-Yau threefold of $\hat X_4$ is precisely $Z_3$.  This shows
that the geometric moduli of $Z_3$ are correctly embedded in $\hat X_4$.  The
polynomials $p_\pm$ encode the perturbative bundles, and the explicit form
\eqref{def-p0p+p-} shows that one has a trivial $SU(1)\times SU(1)$ bundle. This
fact can also be directly checked by analyzing the polyhedron of $\hat X_4$
using the methods of \cite{Bershadsky:1996nh,Candelas:1996su}.  One shows explicitly that over each
divisor $k_i=0$ in $B_3$ a full $E_8$ gauge group is realized. Since the full
$E_8 \times E_8$ gauge symmetry is preserved we are precisely in the
situation of section \ref{F_blowup}, where we recalled from
ref.~\cite{Diaconescu:1999it} that a smooth $\hat X_4$ contains a blow-up
corresponding to a heterotic five-brane.  We will now check that this allows us
to identify the brane moduli in the duality.

Let us now make contact to section \ref{F_blowup}.  To make the perturbative
$E_8\times E_8$ gauge group visible in the Weierstrass equation
\eqref{eqn:weierstrass-x4}, we have to include new coordinates $(\tilde
k_1,\tilde k_2)$ replacing $(k_1,k_2)$. This can be again understood by
analyzing the toric data using the methods of
\cite{Candelas:1996su,Bershadsky:1996nh}.  We denote by $(3,2,\vec{\mu})$ the
toric coordinates of the divisor corresponding to $\tilde k_1$ in the
Weierstrass model. Then the resolved $E_8$ singularity corresponds to the
points\footnote{Note that we have chosen the vertices in the $\P_{1,2,3}[6]$ to
be $(-1,0),(0,-1),(3,2)$ to match the discussion in
refs.~\cite{Bershadsky:1996nh,Candelas:1996su}. However, if one explicitly
analyses the polyhedron of $\hat X_4$ one finds that one has to apply a
$Gl(2,\Z)$ transformation to find a perfect match. This is due to the fact that
$\hat X_4$, in comparison to its mirror Mir($\hat X_4$), actually contains the
dual torus as elliptic fiber.}
\begin{eqnarray}
	&(3,2,n \vec{\mu}),\ n=1,...,6\quad ,\quad (2,1,n \vec{\mu}),\ n=1,...,4\ ,&\\
	\nn&(1,1,n \vec{\mu}),\ n=1,2,3\quad ,\quad (1,0,n \vec{\mu}),\ n=1,2\quad,\quad (0,0,\vec{\mu})&
\end{eqnarray}
While $(3,2,6 \vec{\mu})$ corresponding to $k_1$ is a vertex of the
polyhedron, $(3,2,\vec{\mu})$ corresponding to $\tilde k_1$ is an inner point.
Using the inner point for $\tilde k_1$, the Weierstrass form $\mu_4$ changes
slightly, while the polynomials $p_0,p_+$ and $p_-$ can still be identified in
the stable degeneration limit.  To determine $g_5$ in \eqref{eqn:brane-g5}, we
compute $g$ of the Weierstrass form in a local patch where $\tilde k_2=1$
\begin{equation}
	g=\tilde k_1^5 \left(b_1 u_1^{18}+b_2 u_1^6 u_2^6 u_3^6+ \tilde k_1
\left(a_1 u_1^{18}+a_2 u_2^{18}+\dots\right)\right)\ .
\end{equation}
The dots contain only terms of order zero or higher in $\tilde k_1$.
Comparing this with \eqref{f'g'}, $g_5$ is given by
\begin{equation}
	g_5=b_1 u_1^{18}+b_2 u_1^6 u_2^6 u_3^6.
	\label{eqn:exp2-g5}
\end{equation}
This identifies $\{g_5=0\}$ with the curve of the five-brane in the base $B_2$
of $Z_3$ and is in accord with \eqref{eqn:exp2-brane-heterotic}.  One concludes
that $\hat X_4$ is indeed a correct fourfold associated to $Z_3$ with the given
five-brane.  As we can see from \eqref{eqn:exp2-g5}, the five-brane has one
modulus.  If we compare $g_5$ with $p_+$, we see that $p_+=v\tilde{z}^6g_5$.  This
nicely fits with the bundle description.  In our configuration, $p_+$ and $p_-$
should describe $SU(1)$-bundle since we have the full unbroken perturbative
$E_8\times E_8$-bundle as described above.  The $SU(1)$-bundles do not have any
moduli, such that the moduli space corresponds to just one
point~\cite{Friedman:1997yq}.  In the explicit discussion of the Weierstrass
form in our setting, $p_+$ has one modulus which corresponds to the modulus of
the five-brane. Note that the Calabi-Yau fourfold $\hat X_4$ is already blown up
along the curve $\tilde k_1=g_5=0$ in the base of $\hat X_4$. This blow-up can
be equivalently described as a complete intersection as we discussed in the
previous sections. A simple example of such a construction was presented in
section \ref{non-CYblowup}.

Finally, we consider the computation of the flux superpotential. Here, we do not
need to recall all the details, since the superpotential for this configuration
was already studied in ref.~\cite{Grimm:2009ef}.  The different triangulations
of Mir$(\hat X_4)$ correspond to different five-brane configurations. The
four-form flux, for one five-brane configurations, was shown to be given in the
basis elements
\beq  \label{expl_flux}
  \hat \gamma_1^{(2)} = \tfrac{1}{2} \theta_4 (\theta_1 + \theta_3)
\Omega_4 |_{\underline z=0} \ , \qquad
  \hat \gamma_1^{(2)} = \tfrac{1}{7} \theta_2 (\theta_2-2 \theta_1 + 6
\theta_4 - \theta_3)\Omega_4 |_{\underline z=0}\ ,
\eeq
where the $\theta_i = z_i \frac{d}{dz_i}$ are the logarithmic derivatives as introduced in \eqref{def-cR}.
The moduli $z_1,z_2$ can be identified as the deformations of the complex
structure of the heterotic threefold $Z_3$, while $z_3$ corresponds to the
deformation of the heterotic five-brane.\footnote{The deformation $z_4$
describes the change in $p_-$.} A non-trivial check of this identification was
already provided in \cite{Grimm:2009ef}, where it was shown that the F-theory
flux superpotential in the directions \eqref{expl_flux} matches with the
superpotential for a five-brane configuration in a local Calabi-Yau threefold
obtained by decompactifying $Z_3$. This non-compact five-brane can be described
by a point on a Riemann surface in the base $B_2$ of $Z_3$. Using heterotic
F-theory duality as in section \ref{F-theory_sec}, one can now argue that the
flux \eqref{expl_flux} actually describes a compact heterotic five-brane
setup.

\section{Conclusion}

In this work we have studied the dynamics of heterotic five-branes 
using the duality between the heterotic string and F-theory compactifications. In particular,
we have exploited the fact that five-branes wrapped on the base of an 
elliptically fibered Calabi-Yau threefold $Z_3$ map under duality into the 
geometry of the F-theory Calabi-Yau fourfold $X_4$. This implies that  the heterotic 
five-brane superpotential has to be identified with a F-theory flux 
superpotential. On the heterotic side the five-brane superpotential 
is given by a chain integral $\int_\Gamma \Omega$ over the holomorphic three-form of $Z_3$. 
Upon identifying the F-theory four-form flux which corresponds to 
this three-chain $\Gamma$, the determination of the superpotential 
becomes a tractable task \cite{Grimm:2009ef}. This is due to the 
fact that the deformation moduli of the five-brane are mapped to 
complex structure deformations of the dual Calabi-Yau fourfold $X_4$. 
Their dynamics is then captured by the periods of the 
holomorphic four-form on $X_4$.

The construction of the F-theory fourfold dual to a five-brane has 
been argued to involve a blow-up of the five-brane curve
\cite{Rajesh:1998ik,Berglund:1998ej}. 
We have provided further evidence for this proposal by noting that
this blow-up can also be performed in the heterotic 
Calabi-Yau threefold. Following our discussion in ref.~\cite{Grimm:2008dq},
the deformation moduli of the five-brane curve become  
new complex structure deformation of the blown-up K\"ahler
threefold. This space is no longer Calabi-Yau and the vanishing of $\hat{\Omega}$ implies,
that the heterotic flux naturally maps to the relative cohomology of $\hat{Z}_3$. 
This allows for an equal treatment of the different parts of the superpotential by expressing the complete
heterotic flux supporting both the five-brane and flux superpotential as derivatives of $\hat{\Omega}$ with 
respect to the complex structure of $\hat{Z}_3$.
Finally, we were able to explicitly show that
there exists a natural map of this non-Calabi-Yau threefold to the 
F-theory Calabi-Yau fourfold. In an upcoming publication \cite{inprogress:GKK} such maps from a more general class 
of non-Calabi-Yau threefolds to Calabi-Yau fourfolds is constructed and verified by explicit computations on both geometries. 

By the identification of the fourfold variables \eqref{flatfourfold}  with the
heterotic variables in the superpotentials $W_{G_4}(\underline{t})=
W_{\text{het}}(\underline{t}^c,\underline{t}^g,\underline{t}^o)$, the integral
structure \eqref{Li2} of the fourfold symplectic invariants at large
volume~\cite{Greene:1993vm,Klemm:2007in}
is now inherited to the heterotic superpotentials in geometric
phases of their parameter spaces.  For the superpotential from  five-branes wrapped
on a curve ${\cal C}$  this matches naturally the disk multi-covering formula
of~\cite{Ooguri:1999bv}, since this part is mapped by mirror symmetry to disk
instantons ending on special Lagrangians $L$ mirror dual to ${\cal C}$. Similar
for the heterotic flux superpotential it matches the expectations from the
rational curve counting on threefolds as encoded in the period of the
threefold. For the Chern-Simons part of the potential we obtain by our
construction integer geometric invariants for the gauge bundles on Calabi-Yau
threefolds whose precise relation to Donaldson-Thomas invariants is an
interesting subject of research. Finally the Picard-Fuchs system of the
fourfold allows to analytically continue the superpotential away from the
geometric phases of the open/gauge and closed moduli space into the interior of
this moduli space and to find the correct open and closed flat coordinates in
these regions, see e.g.~for the orbifold points~\cite{Bouchard:2008gu}.

Let us point out some applications of our results. Firstly, the computation of
the superpotential is crucial in the study of moduli stabilization. The
F-theory fourfold setup provides powerful tools to determine heterotic vacua
in which five-brane and bundle moduli are stabilized. As the F-theory flux
superpotential can be determined at an arbitrary point in the moduli space, one
is able to study a landscape of heterotic vacua with five-branes and
gauge-bundle configurations far inside the moduli space.  Since the F-theory
K\"ahler potential for the complex structure moduli of the Calabi-Yau fourfold
is computable as a function of the periods of the holomorphic four-form, one
also expects to determine the kinetic terms of the five-brane and bundle moduli
at different points in the moduli space. It will be an interesting task to
explicitly determine the heterotic K\"ahler potential close to singular
configurations and to search for interesting supersymmetric and non-supersymmetric
vacua in analogy to the Type IIB analysis~\cite{Polchinski:1995sm,Douglas:2006es,Denef:2008wq}.

A second application will be the study of the heterotic compactifications which
are dual to phenomenologically appealing F-theory vacua. Recently, in
refs.~\cite{Blumenhagen:2009yv}, a promising class of Calabi-Yau
fourfolds for GUT model building was constructed by blowing up singular curves
in the base of an elliptic fourfold.  The geometries were explicitly realized
as complete intersections in a toric ambient space. Remarkably, these manifolds
share various properties with the geometries constructed in this work.  To
explore this relation and the use of heterotic/F-theory duality in more detail
will be an interesting and important task \cite{workinprogress}.

\subsection*{Acknowledgments}

We gratefully acknowledge discussions with Ralph Blumenhagen, Johannes Walcher
and especially Timo Weigand.  T.G.~would like to thank the MPI Munich for
hospitality.  This work was supported in parts by the European
Union 6th framework program MRTN-CT-2004-503069 ``Quest for unification'',
MRTN-CT-2004-005104 ``ForcesUniverse'', MRTN-CT-2006-035863 ``UniverseNet'',
SFB-Transregio 33 ``The Dark Universe'' by the DFG.  The work of T.-W.H.\ and
D.K.\ is supported by the German Excellence Initiative via the graduate school
``Bonn Cologne Graduate School". The work of D.K.\ is supported by a
scholarship of the ``Deutsche Telekom Stiftung".

\end{document}